\documentclass[aps,pra,superscriptaddress,twocolumn]{revtex4-2}
\usepackage{xcolor}
\usepackage{natbib}
\usepackage{amsmath}
\usepackage{amssymb}
\usepackage{amsthm}
\usepackage{nicefrac}
\usepackage{graphicx}
\usepackage{bm}
\usepackage{url}
\usepackage{physics}
\usepackage{subcaption}

\renewcommand{\vec}[1]{\bm{\mathrm{#1}}}
\renewcommand{\cos}[1]{\text{cos}\left(#1\right)}
\renewcommand{\sin}[1]{\text{sin}\left(#1\right)}

\newcommand{\vecop}[1]{\hat{\mathbf{#1}}}
\newcommand{\bracket}[1]{\langle #1 \rangle}

\newcommand{\TT}{\mathbb{T}}

\newcommand{\VV}{\mathbb{V}}

\newcommand{\II}{\mathbb{I}}

\newcommand{\PP}{\mathbb{P}}
\newcommand{\Gvac}{\mathbb{G}_{0}}
\newcommand{\Gsca}{\mathbb{G}_{\mathrm{sca}}}

\begin{document}


\title{Trace expressions and associated limits for equilibrium Casimir torque}

\author{Benjamin Strekha}
\affiliation{Department of Electrical and Computer Engineering, Princeton University, Princeton, New Jersey 08544, USA}

\author{Matthias Kr\"{u}ger}
\affiliation{Institute for Theoretical Physics, Georg-August-Universit\"{a}t G\"{o}ttingen, 37073 G\"{o}ttingen, Germany}

\author{Alejandro W. Rodriguez}
\affiliation{Department of Electrical and Computer Engineering, Princeton University, Princeton, New Jersey 08544, USA}


\begin{abstract}
We exploit fluctuational electrodynamics to present trace expressions for the torque experienced by arbitrary objects in a passive, nonabsorbing, rotationally invariant background environment. 
We present trace expressions for equilibrium Casimir torque which complement recently derived nonequilibrium torque expressions and explicate their relation to the Casimir free energy.
We then use the derived trace expressions to calculate, via Lagrange duality, semianalytic structure-agnostic bounds on the Casimir torque
between an anisotropic (reciprocal or nonreciprocal) dipolar particle and
a macroscopic body composed of a local isotropic electric susceptibility, separated by vacuum.
\end{abstract}

\maketitle



Fluctuation phenomena have been successfully
studied and measured in the past few decades~\cite{klimchitskaya2009casimir,woods2016materials}.
Experiments involving a sphere-plate setup have measured radiative heat transfer~\cite{rousseau2009radiative,cuevas2018radiative}, an attractive Casimir force~\cite{lamoreaux1997demonstration,mohideen1998precision}, and a long-range repulsive Casimir force~\cite{munday2009measured}.
Casimir torque, on the other hand, has been less studied up to date.
In anisotropic media or systems exhibiting chirality, thermal
fluctuations can also cause objects to exchange net angular momentum
with their environments or other nearby objects, resulting in a predicted net
torque~\cite{kats1971nonisotropic,parsegian1972dielectric,chinmay2021forcetorque,guo2021single,gao2021thermal,strekha2022tracenoneq}.
Several theoretical proposals to detect
the Casimir torque include birefringent plates~\cite{munday2005torque}, anisotropic nanostructures
~\cite{rodrigues2006vacuum,guerout2015casimir}, and liquid crystals~\cite{somers2015rotation}. 
The challenge in each system for detection stems from
the weakness of the effect and the requirement of ultra-sensitive measurements;
a larger effect requires the ability to place objects at a small separation.
The prediction of a Casimir torque was eventually confirmed experimentally in 2018
between a liquid crystal and a solid birefringent crystal where,
to ensure that the two surfaces were parallel (a problem less relevant for sphere-plate setups), the vacuum gap was replaced by an isotropic
material to act as a spacer layer~\cite{somers2018measurement}.
The measurement of Casimir torque between two objects (a nanorod levitated by a linearly polarized optical
tweezer near a birefringent plate) with a vacuum
gap was proposed in Ref.~\cite{xu2017detecting} but has not yet been realized, although recent experiments~\cite{ahn2018optically,ahn2020ultrasensitive} with an optically levitated nanodumbbell or nanosphere in vacuum demonstrated a torque detection sensitivity on the order of $10^{-27}~\textrm{N}\cdot \textrm{m}\cdot \textrm{Hz}^{-1/2}$, demonstrating progress towards detecting
a Casimir torque in a setup involving a dipolar particle near a macroscopic body, separated by vacuum.

In calculating equilibrium Casimir forces on a rigid body, it is common to start with a Casimir free energy 
and then take a negative derivative with respect to the position of the body to get a force~\cite{rahi2009scattering,emig2007casimir,gelbwaser2022equilibrium}.
The equilibrium Casimir torque is expected to be a rotational derivative of the Casimir free energy,
and indeed was calculated in the above cited works from the Casimir free energy.
In this article, we start from the general Lorentz force law and then use the mathematical 
framework of fluctuational electrodynamics~\cite{rytov1989principles,otey2014fluctuational} and
scattering theory~\cite{rahi2009scattering} to derive trace
expressions for the thermal Casimir forces and torque experienced by a set of
objects in thermal equilibrium which elucidate in a unified framework the precise relationship of
force and torque to the Casimir free energy.
In particular, the torque is confirmed to be a rotational derivative of the Casimir free energy but, since photons are
spin-1 particles, the rotation is an angular rotation 
of the center of mass of the body plus a rotation of the vector components of the scattering operator.

As an application of the derived torque expressions, we calculate bounds on the Casimir torque on an anisotropic dipolar particle from neighboring objects.
There is the interesting question of whether Casimir torques must be weak.
As the ability to manipulate mechanical devices of increasingly smaller scales continues to increase, so too will increase the interest in
exploiting fluctuation phenomena such as laser shot noise and the Casimir effect as a mechanism of
control in micromachines~\cite{ding2022universal,van2021sub,stickler2021quantum}.
The natural question of which geometry leads to maximum torque can be probed via large-scale optimization but cannot,
in general, provide guarantees of global optimality~\cite{molesky2018inverse,christiansen2021inverse}.
Further understanding, e.g. quantitative bounds and scaling behavior, can be gained by applying a recent
framework based on Lagrange duality to compute shape-independent bounds~\cite{T_operator_bounds_angle_integrated,global_T_operator_bounds,chao2022physical,strekha2022tracenoneq,venkataram2020fundamentalCPforce,strekhachao2023minldos}. 
In particular, Ref.~\cite{venkataram2020fundamentalCPforce} presents bounds on the surface-perpendicular Casimir force on a dipole above a half-space design domain and the authors remark that similar methods can be used for bounds on Casimir torques by taking a derivative of the dipolar basis functions with respect to a rotation angle.
Naive manipulations and rotational derivatives will yield incorrect results by missing ``spin'' contributions.
Our derived expressions with an explicit appearance of a total angular momentum operator $\hat{\mathbf{J}}$ allow for a more lucid analysis of torque phenomena.

\section{Equilibrium Casimir Effects}

As explained in detail in Refs.~\cite{kruger_trace_formulae_for_nonequilibrium,strekha2022tracenoneq,zhang2022microscopic,wang2023transport,wang2023photon}, starting from the Lorentz force law one can show that the 
thermally averaged $(\langle\dots\rangle_{T})$ rate of absorption associated with an observable $\hat{\Theta}$ is 
given by
\begin{align}
    \langle\hat{\Theta}\rangle_{T} &= -\textrm{Im}\int_{-\infty}^{\infty} \mathrm{d}\omega
    \frac{1}{\hbar\omega^{2}\mu_{0}} \text{Tr}|_{V_{\textrm{body}}}[\hat{\Theta} \mathbb{C}\mathbb{G}_{0}^{-1\dagger}],
    \label{eq:torquetracewithC} 
\end{align}
where substituting $\hbar\omega\mathbb{I}, \hat{\mathbf{p}}, \hat{\mathbf{J}},$ for $\hat{\Theta}$ above yields the absorbed power, force, and torque on the body, respectively.
Here, $\hat{\mathbf{p}} \equiv -i\hbar\nabla$ is the linear momentum operator, $\hat{\mathbf{J}} \equiv \hat{\mathbf{L}} + \hat{\mathbf{S}}$ is the total angular momentum operator, $\hat{\mathbf{L}} \equiv \mathbf{r}\times\hat{\mathbf{p}} = -i\hbar\mathbf{r}\times\nabla$, and
\begin{align}
    \vecop{S} \equiv -i\hbar
    \Bigg\{ 
    \begin{bmatrix}
        0 & 0 & 0 \\
        0 & 0 & 1 \\
        0 & -1 & 0 
    \end{bmatrix},
    \begin{bmatrix}
        0 & 0 & -1 \\
        0 & 0 & 0 \\
        1 & 0 & 0 
    \end{bmatrix},
    \begin{bmatrix}
        0 & 1 & 0 \\
        -1 & 0 & 0 \\
        0 & 0 & 0 
    \end{bmatrix}
    \Bigg\},
\end{align}
are orbital and spin angular momentum operators~\cite{khersonskii1988quantum},
respectively, defined in the Cartesian
basis and compactly summarized by $(\hat{S}_{a})_{bc} = -i\hbar \epsilon_{abc}$. 
The notation $|_{V_{\textrm{body}}}$ denotes
that the outer-most indices of the operator are traced over positions
in the body, while all others are over all space, and a trace over the vector components of the Dyadics. That is,
$\text{Tr}|_{V}[\mathbb{A}\mathbb{B}] = \sum_{ab}\int_{\mathbf{r}\in V}\int_{\mathbf{s}\in \mathbb{R}^{3}} \mathrm{d}^{3}\mathbf{r}\mathrm{d}^{3}\mathbf{s} \mathbb{A}_{ab}(\mathbf{r}, \mathbf{s})\mathbb{B}_{ba}(\mathbf{s}, \mathbf{r}).$
In other words, $\text{Tr}|_{V}[(\cdots)] = \text{Tr}[\mathbb{P}(V)(\cdots)]$ where $\mathbb{P}(V)$ is a projection operator into the spatial volume $V.$
The operator $\mathbb{G}_{0}$ represents the background Green's function, which in
vacuum satisfies $\left[ \nabla\times\nabla\times -
  \frac{\omega^{2}}{c^{2}}\mathbb{I}\right]\mathbb{G}_{0}(\mathbf{r},
\mathbf{r}') = \mathbb{I}\delta^{(3)}(\mathbf{r} - \mathbf{r}').$ For systems in thermal
equilibrium the field--field
correlations satisfy $\mathbb{C}_{ij}^{\textrm{eq}}(T, \omega,\omega'; \mathbf{r},
\mathbf{r}') \equiv \langle E_{i}(\mathbf{r},
\omega)E_{j}^{*}(\mathbf{r}', \omega ') \rangle_{T} =
\frac{\hbar\omega^{2}}{2\pi c^{2}\epsilon_{0}} \coth\left(
\frac{\hbar\omega}{2k_{B}T} \right) \delta(\omega - \omega')
\mathbb{G}^{\mathsf{A}}_{ij}(\omega; \mathbf{r}, \mathbf{r}')$, where $\mathbb{G}$ is the Green's function of the system, defined by
$\left[ \nabla\times\nabla\times - \mathbb{V} -
  \frac{\omega^{2}}{c^{2}}\mathbb{I}\right]\mathbb{G}(\mathbf{r},
\mathbf{r}') = \mathbb{I}\delta^{(3)}(\mathbf{r} - \mathbf{r}')$ where
$\mathbb{V} = \frac{\omega^{2}}{c^{2}}(\mathbb{\epsilon} - \mathbb{I})
+ \nabla\times(\mathbb{I} - \mathbb{\mu}^{-1})\nabla\times$ is the
potential or generalized susceptibility introduced by the
objects~\cite{kruger_nonequilbrium_fluctuations_review}. The superscript $\mathsf{A}$ on an operator $\hat{\Theta}$ denotes a Hermitian operator which
contains the anti-Hermitian part of $\hat{\Theta}$, defined by $\hat{\Theta}^{\mathsf{A}} \equiv
\frac{1}{2i}(\hat{\Theta} - \hat{\Theta}^{\dagger}),$ where $\dagger$ denotes
conjugate transpose. In our notation,
$\hat{\Theta}_{ab}^{\dagger}(\mathbf{x}, \mathbf{y}) =
\hat{\Theta}_{ba}(\mathbf{y}, \mathbf{x})^{*}$, treating the vector
component and spatial coordinate as an index pair.

For a single body, the Green's function satisfies $\mathbb{G} = \mathbb{G}_{0} + \mathbb{G}_{0}\mathbb{T}\mathbb{G}_{0}$ where we have introduced the scattering
$\mathbb{T}$ operator which transforms incident fields into
induced currents in the body and is
formally defined by the relation $\mathbb{T} = \mathbb{V}(\mathbb{I} -
\mathbb{G}_{0}\mathbb{V})^{-1}$ ~\cite{kruger_trace_formulae_for_nonequilibrium}. Since $\mathbb{T}$ vanishes unless both spatial arguments are within the body, the integral can be extended over all space to get a trace expression
\begin{align}
    \langle\hat{\Theta}\rangle_{T}^{\textrm{eq}} =
    \text{Re}\int_{-\infty}^{\infty} \frac{\mathrm{d}\omega}{4\pi} \coth\left(\frac{\hbar\omega}{2k_{B}T} \right)
    \text{Tr}[\hat{\Theta}(\mathbb{G}_{0}\mathbb{T} - \mathbb{G}_{0}^{\dagger}\mathbb{T}^{\dagger})].
\end{align}
However, $\text{Re}\text{Tr}[\hat{\Theta}(\mathbb{G}_{0}\mathbb{T} - \mathbb{G}_{0}^{\dagger}\mathbb{T}^{\dagger})] = 0$ independent of $\mathbb{V}$ (reciprocal or nonreciprocal) so the net power absorption, force, and torque on an isolated object in equilibrium is identically zero, as expected.

Consider next the case of two or more bodies.
The Green's function is found by starting with
object 1 in isolation with $\mathbb{G}_{1} = (1 + \mathbb{G}_{0}\mathbb{T}_{1})\mathbb{G}_{0}$ and then inserting the rest of the objects to get $\mathbb{G} = (1 + \mathbb{G}_{0}\mathbb{T}_{\bar{1}})\frac{1}{1 - \mathbb{G}_{0}\mathbb{T}_{1}\mathbb{G}_{0}\mathbb{T}_{\bar{1}}}(1 + \mathbb{G}_{0}\mathbb{T}_{1})\mathbb{G}_{0}$, where $\mathbb{T}_{\bar{1}}$ denotes the scattering operator of all objects excluding object 1~\cite{kruger_trace_formulae_for_nonequilibrium}.
Changing the order of object insertion implies that the same equation with the indices swapped $1 \leftrightarrow \bar{1}$ holds as well.
Using these expressions for the Green's function in the equilibrium field--field correlation Dyadic $\mathbb{C}^{\textrm{eq}}$ in Eq.~\eqref{eq:torquetracewithC} we find that the thermally averaged torques on the objects are (intermediate steps are provided in the Appendix)
\begin{multline}
    \boldsymbol{\tau}^{(\alpha,\textrm{eq})}(T) =
    \text{Re}\frac{1}{\pi}\int_{0}^{\infty}\mathrm{d}\omega [n(\omega, T) + \frac{1}{2}] \\
    \times\text{Tr}\bigg[\frac{1}{1 - \mathbb{G}_{0}\mathbb{T}_{\alpha}\mathbb{G}_{0}\mathbb{T}_{\bar{\alpha}}}\mathbb{G}_{0}
    (\mathbb{T}_{\alpha}\vecop{J} - \vecop{J}\mathbb{T}_{\alpha})\mathbb{G}_{0}\mathbb{T}_{\bar{\alpha}}\bigg]
    \label{eq:taualphaeqalmostln}
\end{multline}
where $n(\omega,T) = \frac{1}{\exp(\frac{\hbar\omega}{ k_{B}T}) - 1}$
is the Bose--Einstein distribution function. The Tr symbol denotes a trace over the complete set of indices of the
enclosed operators (for example, over all positions and polarization
indices of the dipole sources).
The switch from
$\text{Tr}|_{V_{\alpha}}$ to $\text{Tr}$ is possible since
$\mathbb{C}^{\textrm{eq}}\mathbb{G}_{0}^{-1\dagger}$ has a $\mathbb{T}_{\alpha}$
or $\mathbb{T}^{\dagger}_{\alpha}$ with $\alpha = 1,\bar{1}$ as the left-most or right-most term in the
expansion (App.~\ref{app:traceexpressionsderviation}).
Since $\mathbb{T}_{\alpha}$ vanishes if at least one of the spatial arguments is outside the volume $V_{\alpha}$ of body $\alpha$,
one can extend the spatial integration to be over the entire space, resulting in a trace expression.
The above equations satisfy $\boldsymbol{\tau}^{(1,\textrm{eq})} = -\boldsymbol{\tau}^{(\bar{1},\textrm{eq})}$ 
(by the cyclicity of the trace along with resumming of $\frac{1}{1 - \mathbb{G}_{0}\mathbb{T}_{\bar{1}}\mathbb{G}_{0}\mathbb{T}_{1}}$), 
as expected since the angular momentum transfer to the far-field should cancel and 
detailed balance should hold.
For the case of reciprocal materials, the expressions can be further simplified to
\begin{multline}
    \boldsymbol{\tau}^{(\alpha,\textrm{eq})}(T) = \text{Re}\frac{2}{\pi} \int_{0}^{\infty} \mathrm{d}\omega [n(\omega, T) + \frac{1}{2}] \\ \times\text{Tr}[
    \vecop{J}\mathbb{G}_{0}\mathbb{T}_{\bar{\alpha}}\mathbb{G}_{0}\mathbb{T}_{\alpha}
    \frac{1}{1-\mathbb{G}_{0}\mathbb{T}_{\bar{\alpha}}\mathbb{G}_{0}\mathbb{T}_{\alpha}}
    ], \label{eq:taualphaeq}
\end{multline}
The expressions for the equilibrium forces are identical but with 
$\hat{\mathbf{J}} \to \hat{\mathbf{p}}$, corresponding to a change in observable from rate of angular momentum
absorption to rate of linear momentum absorption.

Equation~\eqref{eq:taualphaeqalmostln} elucidates
the relationship of the torque (and force) with the Casimir free energy $\mathcal{F}$, defined as
\begin{align}
    \mathcal{F} &\equiv -\frac{\hbar}{\pi}\int_{0}^{\infty}\mathrm{d}\omega [n(\omega, T) + \frac{1}{2}] \text{Im}\text{Tr}\bigg[\ln(1 - \mathbb{G}_{0}\mathbb{T}_{1}\mathbb{G}_{0}\mathbb{T}_{\bar{1}})\bigg], \\
    &=
    -\frac{\hbar}{\pi}\int_{0}^{\infty}\mathrm{d}\omega [n(\omega, T) + \frac{1}{2}] \text{Im}\ln[\det(1 - \mathbb{G}_{0}\mathbb{T}_{1}\mathbb{G}_{0}\mathbb{T}_{\bar{1}})],
\end{align}
more directly than Eq.~\eqref{eq:taualphaeq}.
Viewing the free energy $\mathcal{F}(\mathbb{T}_{\alpha})$ as a function of $\mathbb{T}_{\alpha} = \mathbb{V}_{\alpha}(\mathbb{I} - \mathbb{G}_{0}\mathbb{V}_{\alpha})^{-1}$, the equilibrium force and torque can be written as
\begin{align}
    \mathbf{F}^{(\alpha,\textrm{eq})} &=
    -\lim_{\Delta\mathbf{x} \to 0}\frac{\mathcal{F}(\exp(-\frac{i\Delta\mathbf{x}\cdot\hat{\mathbf{p}}}{\hbar})\mathbb{T}_{\alpha}\exp(\frac{i\Delta\mathbf{x}\cdot\hat{\mathbf{p}}}{\hbar})) - \mathcal{F}(\mathbb{T}_{\alpha})}{\Delta\mathbf{x}},
\end{align}
\begin{align}
    \boldsymbol{\tau}^{(\alpha,\textrm{eq})}
    &= -\lim_{\Delta\bm{\phi} \to 0} \frac{\mathcal{F}(\exp(-\frac{i\Delta\bm{\phi}\cdot\hat{\mathbf{J}}}{\hbar})\mathbb{T}_{\alpha}\exp(\frac{i\Delta\bm{\phi}\cdot\hat{\mathbf{J}}}{\hbar})) - \mathcal{F}(\mathbb{T}_{\alpha})}{\Delta\bm{\phi}},
    \label{eq:torqueJrotation}
\end{align}
as can be verified by taking the limit directly and recovering Eq.~\eqref{eq:taualphaeqalmostln}. Thus, the equilibrium force and torque on object $\alpha$ are changes in free energy due to a rigid translation and rotation of the scattering operator $\mathbb{T}_{\alpha}$ where, because photons are spin-1 particles, the relevant rotation operator $\exp(-\frac{i\Delta\boldsymbol{\phi}\cdot\hat{\mathbf{J}}}{\hbar})$ depends on $\hat{\mathbf{J}}$ and not $\hat{\mathbf{L}}.$ 

Assuming rigid bodies, the equilibrium Casimir force can be written more explicitly as a negative gradient of a 
free energy $\mathcal{F}$ (Appendix A of Ref.~\cite{kruger_trace_formulae_for_nonequilibrium}).
Concisely, assuming rigid bodies, then $\nabla_{\mathcal{O}_{1}}\mathbb{T}_{1} = \nabla \mathbb{T}_{1} - \mathbb{T}_{1}\nabla = \frac{i}{\hbar}\hat{\mathbf{p}}\mathbb{T}_{1} -\frac{i}{\hbar}\mathbb{T}_{1}\hat{\mathbf{p}}$ where
$\nabla_{\mathcal{O}_{1}}$ is a derivative with respect to the center of mass coordinate of object 1 so that
\begin{align}
    \mathbf{F}^{(1,\textrm{eq})}(T) &= \text{Im}\frac{\hbar}{\pi}\int_{0}^{\infty}\mathrm{d}\omega [n(\omega, T) + \frac{1}{2}]  \nonumber \\
    &\qquad\qquad\times\text{Tr}\nabla_{\mathcal{O}_{1}}\ln\bigg[1 - \mathbb{G}_{0}\mathbb{T}_{1}\mathbb{G}_{0}\mathbb{T}_{\bar{1}}\bigg], \\
    &= -\nabla_{\mathcal{O}_{1}} \mathcal{F}
\end{align}
where the last line follows since $\nabla_{\mathcal{O}_{1}}$ can be brought outside the trace. When calculating equilibrium torque, care must be taken to check if the torque is an angular
derivative of the same free energy $\mathcal{F}.$ In trying analogous steps for torque starting
from Eq.~\eqref{eq:taualphaeqalmostln} we find that the trace expressions cannot, in general, 
be rewritten as angular derivatives with respect to the center of mass coordinates of the rigid body, $-\mathbf{r}_{\mathcal{O}_{1}}\times\nabla_{\mathcal{O}_{1}}\mathcal{F}$, due to the spin terms.

Some intuition can be gleaned from a dilute limit. In the Rytov formalism for fluctuation
electrodynamics~\cite{rytov1989principles}, the object is viewed as a collection of free dipoles which undergo fluctuations.
In thermal equilibrium, the fluctuation-dissipation theorem states that the fluctuations are proportional to the polarizability tensor~\cite{novotny2012principles,kruger_nonequilbrium_fluctuations_review}. 
Consider two objects (viewed as two collections of fluctuating free dipoles) separated by a gap, and call the free energy for a given initial arrangement $\mathcal{F}_{i}.$ Then calculate the free energy when one of them is geometrically rotated by $\Delta \phi$ about the $z$-axis but the polarizability/susceptibility tensor at the rotated spatial point is unchanged and call that final energy $\mathcal{F}_{f,L}.$ That is, rotate the position of each dipole but not its orientation. Likewise, starting from the initial arrangement, keep the position of each dipole fixed but now instead rotate the orientation by $\Delta\phi$ around the $z$-axis and call the resulting free energy $\mathcal{F}_{f,S}.$ The torque on the object is given by $\lim_{\Delta\phi \to 0} - \frac{(\mathcal{F}_{f,L} - \mathcal{F}_{i}) + (\mathcal{F}_{f,S} - \mathcal{F}_{i})}{\Delta\phi}.$ 
Physically, both the spatial location and orientation of each free dipole are rotated during the rotation of an object, which is precisely what Eq.~\eqref{eq:torqueJrotation} states and Eq.~\eqref{eq:taualphaeqalmostln} calculates.

As a word of caution, this simple interpretation breaks down past the dilute limit,
in which case one can no longer view torque as due to a sum of pair-wise interactions
$\boldsymbol{\tau}^{(\alpha,\textrm{eq})} \propto \text{Tr}[
\vecop{J}\mathbb{G}_{0}\mathbb{V}_{\bar{\alpha}}\mathbb{G}_{0}\mathbb{V}_{\alpha}
].$
In particular, if
$\mathbb{V}_{\alpha} \to \exp(-\frac{i\Delta\bm{\phi}\cdot\hat{\mathbf{J}}}{\hbar})\mathbb{V}_{\alpha}\exp(\frac{i\Delta\bm{\phi}\cdot\hat{\mathbf{J}}}{\hbar}),$
then
$\mathbb{T}_{\alpha} \to \exp(-\frac{i\Delta\bm{\phi}\cdot\hat{\mathbf{J}}}{\hbar})\mathbb{T}_{\alpha}\exp(\frac{i\Delta\bm{\phi}\cdot\hat{\mathbf{J}}}{\hbar}),$
as can be seen from the definition $\mathbb{T}_{\alpha} = \mathbb{V}_{\alpha}(1 - \mathbb{G}_{0}\mathbb{V}_{\alpha})^{-1}$ and the fact that $\hat{\mathbf{J}}\mathbb{G}_{0} = \mathbb{G}_{0}\hat{\mathbf{J}}.$ Similar statements do \emph{not} hold for the $\hat{\mathbf{S}},\hat{\mathbf{L}}$ operators since the spin and orbital angular momentum operators do not commute with $\mathbb{G}_{0}.$ For example, the statement that if
$\mathbb{V}_{\alpha} \to \exp(-\frac{i\Delta\bm{\phi}\cdot\hat{\mathbf{S}}}{\hbar})\mathbb{V}_{\alpha}\exp(\frac{i\Delta\bm{\phi}\cdot\hat{\mathbf{S}}}{\hbar}),$
then
$\mathbb{T}_{\alpha} \to \exp(-\frac{i\Delta\bm{\phi}\cdot\hat{\mathbf{S}}}{\hbar})\mathbb{T}_{\alpha}\exp(\frac{i\Delta\bm{\phi}\cdot\hat{\mathbf{S}}}{\hbar})$ is \emph{not} generally true beyond the dilute limit. Instead, one would need to use $\mathbb{T}_{\alpha} \to \mathbb{\tilde{V}}_{\alpha}(1 - \mathbb{G}_{0}\mathbb{\tilde{V}}_{\alpha})^{-1}$ where $\mathbb{\tilde{V}}_{\alpha} \equiv \exp(-\frac{i\Delta\bm{\phi}\cdot\hat{\mathbf{S}}}{\hbar})\mathbb{V}_{\alpha}\exp(\frac{i\Delta\bm{\phi}\cdot\hat{\mathbf{S}}}{\hbar}).$

\section{Bounds on Equilibrium Casimir torque}

In this section, we consider a Wick rotation consistent with a change of variables $\omega = i\xi$ from $\omega$ to $\xi$. Unless stated otherwise, all quantities in this section are taken to implicitly depend on $i\xi.$
As notation, a vector field $\mathbf{v}(\mathbf{x})$ will be denoted as $\ket{\mathbf{v}}$.
At $\omega = i\xi$, although all relevant polarization and field quantities as well as $\mathbb{V}$ and $\mathbb{G}_{0}$
can be defined to be real-valued in position space without loss of generality, we still use the Hermitian inner
product $\bra{\vec{u}}\ket{\vec{v}} = \int\mathrm{d}^{3} x~\vec{u} (\vec{x})^{*} \cdot \vec{v}(\vec{x})$ since the eigenbasis for which calculations are most convenient can still be complex-valued.
An operator $\mathbb{A}(\vec{x}, \vec{x}')$ will be denoted as $\mathbb{A}$, with $\int\mathrm{d}^{3}
x'~\mathbb{A}(\vec{x}, \vec{x}') \cdot \vec{v}(\vec{x}')$ denoted as
$\mathbb{A}\ket{\vec{v}}$.

As described in more detail below, following the procedure laid out in Refs.~\cite{chao2022physical,venkataram2020fundamentalCPforce,global_T_operator_bounds}, and as an exemplary application of Eq.~\eqref{eq:taualphaeqalmostln}, we derive upper and lower bounds on the Casimir torque that may be experienced by a dipolar particle neighboring a structured surface.
The torque bounds necessitate only that the design region $\Omega$ containing the structured object be rotationally invariant about the same axis going through the dipolar particle about which torque is being considered.
The bounds encompass \emph{any} possible structure composed of a homogenous, local, and isotropic electric susceptibility $\chi(i\xi)$ in the design domain $\Omega$.
Concisely, we expand the dipolar response,
including local field effects, along its principal axes as
$\frac{1}{2}(\TT_{\mathrm{dip}}(i\xi) + \TT_{\mathrm{dip}}^{T}(i\xi)) \equiv \sum_{a} \alpha_{a}(i\xi)
\ket{\vec{u}^{(a)}} \bra{\vec{u}^{(a)}}$, where each
$\alpha_{a} (i\xi) > 0$ is a polarizability component, while
$\vec{u}^{(a)}(\vec{x}) = \vec{n}_{a} \delta^{(3)} (\vec{x} -
\vec{R})$ corresponds to a localized basis function at $\vec{R}$ (the dipole location) 
along the $\vec{n}_{a}$ direction.
The chosen design domain $\Omega$ enclosing the second object is such that projection into $\Omega$, denoted $\mathbb{P}(\Omega)$,
commutes with the observable of interest $\hat{\Theta}.$
Our bounds $\Theta \in [\Theta^{-}, \Theta^{+}]$ for $T=0$~K can be written concisely as
\begin{multline} 
  \Theta^{\pm} = \pm \int_{0}^{\infty} \frac{\mathrm{d}\xi}{2\pi} \sum_{a}
  \alpha_{a}
  \sqrt{\frac{\left\langle\vec{u}^{(a)}, \hat{\Theta}^{\dagger}\hat{\Theta}\Gsca(\Omega)
    \vec{u}^{(a)}\right\rangle }{\left\langle
      \vec{u}^{(a)}, \Gsca(\Omega)
      \vec{u}^{(a)}\right\rangle^{-1}}},
      \label{eq:CPThetabounds}
\end{multline}
where $\Gsca(\Omega) = \Gvac
(\chi^{-1} \PP(\Omega) - \PP(\Omega)\Gvac\PP(\Omega))^{-1} \Gvac$ is
the scattering Green's function of the equivalent object formed by
filling the entire domain $\Omega$ with the susceptibility
$\chi$.
The maximization and minimization bounds differ in sign, but not magnitude, as expected on physical grounds if $[\hat{\Theta}, \mathbb{P}(\Omega)]=0.$

\emph{Derivation}--- For concreteness, consider the Casimir torque at zero temperature on a dipolar particle, which can be written as
\begin{multline}
        \boldsymbol{\tau}_{\mathrm{dip}} = \int_{0}^{\infty} \frac{\mathrm{d}\xi}{2\pi} \text{Re}\text{Tr}\bigg[\frac{1}{1 - \mathbb{G}_{0}\mathbb{T}_{\mathrm{dip}}\mathbb{G}_{0}\mathbb{T}}\times \\
        \mathbb{G}_{0}(\mathbb{T}_{\mathrm{dip}}i\vecop{J} - i\vecop{J}\mathbb{T}_{\mathrm{dip}})\mathbb{G}_{0}\mathbb{T}\bigg]
\end{multline}
where $\TT = (\VV^{-1} - \Gvac)^{-1} = (1 - \VV\Gvac)^{-1} \VV$ and
likewise $\TT_{\mathrm{dip}} = (\VV_{\mathrm{dip}}^{-1} - \Gvac)^{-1}
= (1 - \VV_{\mathrm{dip}} \Gvac)^{-1} \VV_{\mathrm{dip}}$.
We group $i$ and $\vecop{J}$ together so that all operators are explicitly real-valued in position space.
To simplify matters, we assume that the dipole is small enough compared to the macroscopic
body that multiple scattering effects between different objects can be neglected, though
multiple scattering within each object cannot. That is, we replace $(1 - \TT_{\mathrm{dip}} \Gvac \TT
\Gvac)^{-1}$ with $1$.
Also, assume for simplicity that $\mathbb{T}$ is symmetric ($\mathbb{V}$ describes a reciprocal material) but that $\mathbb{T}_{\mathrm{dip}}$ can be nonreciprocal. Finally, using the cyclic properties
of the trace, the fact that $\hat{J}^{T}_{k} = -\hat{J}_{k},$ and the assumption that $\mathbb{G}_{0}$ and $\mathbb{T}$ are symmetric yields 
\begin{align}
    \tau_{\mathrm{dip},k}
    &= 
    \int_{0}^{\infty} \frac{\mathrm{d}\xi}{2\pi} 
    \text{Re} \text{Tr}
    \bigg[\mathbb{G}_{0}
    (\mathbb{T}_{\mathrm{dip}} + \mathbb{T}_{\mathrm{dip}}^{T})i\hat{J}_{k}\mathbb{G}_{0}\mathbb{T}\bigg], \\
    &= \int_{0}^{\infty}
    \frac{\mathrm{d}\xi}{2\pi}
    \text{Re}2\sum_{a}
    \alpha_{a}
    \left\langle
    \vec{u}^{(a)}, i\hat{J}_{k}\Gvac\TT\Gvac \vec{u}^{(a)}
    \right\rangle.
\end{align}
The torque on the dipole depends only on the symmetric part of $\mathbb{T}_{\mathrm{dip}}$,
and it is this expression for the torque on the dipole which we will maximize and minimize.


Our derivation of bounds is based on optimization using the principles of Lagrangian duality~\cite{boyd2004convex}.
The loosest such bound only imposes that the optimal scattering operator satisfies the
conservation of power (optical theorem~\cite{jackson1999classical})
over the entire design domain, and not the full scattering equations. Instead of optimizing over $\mathbb{V}$ which has support in $\Omega,$
we relax the problem and instead take as the degree of freedom the induced (polarization) current $|\mathbf{P}^{(a)}\rangle \equiv \mathbb{T}\mathbb{G}_{0}|\mathbf{u}^{(a)}\rangle$ with support only in the design domain $\Omega$.
Since the polarization current is the degree of freedom, the bounds are monotonic with respect to the design domain--- any polarization currents explored in an optimization with support within some smaller domain $\Omega' \subset \Omega$ are also explored in the bigger domain $\Omega.$
Similar techniques have recently been used to derive bounds on deterministic scattering and nonequilibrium electromagnetic phenomena~\cite{global_T_operator_bounds,molesky2020hierarchical,venkataram2020fundamentalCPforce,chao2022physical,amaolo2023canheterostructures,strekhachao2023minldos,angeris2019computational,shim2019fundamental,strekha2022tracenoneq}.
Concretely, the problem we solve is
\begin{subequations}
\begin{multline}
    \max_{\{\vb{P}_{a}(\vb{r}; i\xi)\in\Omega\}} \quad 
    \int_{0}^{\infty}
    \frac{\mathrm{d}\xi}{2\pi}
    2\sum_{a}
    \alpha_{a}
    \text{Re}
    \left\langle
    \vec{u}^{(a)}, \Gvac i\hat{J}_{k}\vec{P}^{(a)}
    \right\rangle
    \label{eq:primal_objective}
\end{multline}
subject to
\begin{multline}
     \text{Re}\bigg[\bra{\mathbf{u}^{(a)}}\Gvac\ket{\vb{P}^{(a)}}
     \\ - \bra{\vb{P}^{(a)}}
     \PP(\Omega)(
     \chi^{-1} - \Gvac
     )\PP(\Omega)
     \ket{\vb{P}^{(a)}}\bigg] = 0
    \label{eq:primal_constraint}
\end{multline}
\label{eq:primal_problem}%
\end{subequations}
for all $a$ and each $\xi \geq 0.$

First, we consider the problem of maximizing $\text{Re}(2\langle \vec{E}^{\mathrm{inc}},
i\hat{\Theta}\vec{P} \rangle).$ For convenience, we define the eigenvalue decomposition of the projection of $\Gvac$ into the
domain $\Omega$ as $\PP(\Omega)\Gvac\PP(\Omega) = -\sum_{\mu}
g_{\mu} \ket{\vec{N}^{(\mu)}} \bra{\vec{N}^{(\mu)}}$, where
$g_{\mu} > 0$, and the eigenvectors are orthonormal:
$\bracket{\vec{N}^{(\mu)}, \vec{N}^{(\nu)}} = \delta_{\mu\nu}$. 
We also assume that $\ket{\vec{N}^{(\mu)}}$ are eigenstates of $\hat{\Theta}$ so that
$\hat{\Theta}\ket{\vec{N}^{(\mu)}} = \theta_{\mu}\ket{\vec{N}^{(\mu)}}.$
For our operators of interest, $\hat{\Theta}^{\dagger} = \hat{\Theta}$ so the eigenvalues $\theta_{\mu}$ are purely real.
Since $\hat{\Theta}$ commutes with isotropic $\mathbb{G}_{0},$ 
the assumption of simultaneous diagonalization therefore puts restrictions on $\mathbb{P}(\Omega)$, i.e. the geometry of the design domain.
We then define the basis expansion coefficients
$e_{\mu} = \bracket{\vec{N}^{(\mu)}, \vec{E}^{\mathrm{inc}}}$ and $p_{\mu} = \bracket{\vec{N}^{(\mu)}, \vec{P}}$.
This leads to a constrained
optimization problem with a Lagrangian given by
\begin{equation}
  L = \sum_{\mu} \bigg[\text{Re}(2e_{\mu}^{*} p_{\mu} i\theta_{\mu}) -
  \lambda \bigg(\text{Re}(e_{\mu}^{*} p_{\mu}) - (\chi^{-1} + g_{\mu})p_{\mu}^{*}p_{\mu}\bigg)\bigg]
\end{equation}
where $\lambda\in\mathbb{R}$ is a Lagrange multiplier.

Carrying out the optimization, we find that $e_{\mu}^{*}i\theta_{\mu} - \lambda[e_{\mu}^{*}/2 - (\chi^{-1} + g_{\mu})p_{\mu}^{*}] = 0$ and $\sum_{\mu}(\text{Re}(e_{\mu}^{*}p_{\mu}) - (\chi^{-1} + g_{\mu})p_{\mu}^{*}p_{\mu}) = 0.$
The first equation gives $p_{\mu} = \frac{e_{\mu}}{\chi^{-1} + g_{\mu}}\big(\frac{1}{2} +\frac{i\theta_{\mu}}{\lambda} \big),$ which can then be plugged into the second equation to get
\begin{align*}
  \lambda &=
  \pm 2\sqrt{
  \bigg(\sum_{\mu}\frac{|e_{\mu}|^{2}\theta_{\mu}^{2}}{\chi^{-1}+g_{\mu}}\bigg)
  \bigg/
  \bigg(\sum_{\mu}\frac{|e_{\mu}|^{2}}{\chi^{-1}+g_{\mu}}\bigg)
  }, \\
  &=\pm 2\sqrt{\frac{\left\langle 
        \hat{\Theta}\vec{E}^{\mathrm{inc}}, 
        (\chi^{-1}
      \PP(\Omega) - \PP(\Omega)\Gvac\PP(\Omega))^{-1}
        \hat{\Theta}\vec{E}^{\mathrm{inc}}
      \right\rangle}{\left\langle\vec{E}^{\mathrm{inc}}, (\chi^{-1}
      \PP(\Omega) - \PP(\Omega)\Gvac\PP(\Omega))^{-1}
      \vec{E}^{\mathrm{inc}}\right\rangle}}.
\end{align*}
The objective has $\frac{\delta^{2} L}{\delta p_{\mu}
  \delta p_{\nu}^{*}} = \lambda (\chi^{-1} + g_{\mu})
\delta_{\mu\nu}$, so the negative value of $\lambda$ gives the maximum
while the positive value gives the minimum.
The stationary point corresponding to $\lambda = 0$, which is a saddle
point, can occur if $\hat{\Theta}\ket{\vec{E}^{\mathrm{inc}}}= 0$.
However, this cannot occur for the incident field radiated by a dipole into the design domain, 
so we ignore this mathematical case going forward.
Consequently, $L \in [L^{-}, L^{+}]$, with
\begin{multline}
  L^{\pm} =\\
    \pm \sqrt{\frac{\left\langle 
      \vec{E}^{\mathrm{inc}}, 
      \hat{\Theta}^{\dagger}(\chi^{-1} \PP(\Omega)
    - \PP(\Omega)\Gvac\PP(\Omega))^{-1} \hat{\Theta}
      \vec{E}^{\mathrm{inc}}
    \right\rangle
    }{\left\langle \vec{E}^{\mathrm{inc}}, (\chi^{-1}
    \PP(\Omega) - \PP(\Omega)\Gvac\PP(\Omega))^{-1}
    \vec{E}^{\mathrm{inc}} \right\rangle^{-1}}}.
\end{multline}

For our problem of interest, Eq.~\eqref{eq:primal_objective}, we set $\ket{\vec{E}^{\mathrm{inc}}} =
\Gvac\ket{\vec{u}^{(a)}}$ and identify $\Gsca(\Omega) = \Gvac
(\chi^{-1} \PP(\Omega) - \PP(\Omega)\Gvac\PP(\Omega))^{-1} \Gvac$ as
the scattering Green's function of the equivalent object formed by
filling the entire domain $\Omega$ with the susceptibility
$\chi$. Since $\alpha_{a}
(i\xi) > 0$, the net upper bound cannot fall above the upper bound
applied to each channel $a$, just as the net lower bound cannot
fall below the per-channel lower bound.
This argument also applies to each $\xi$ in the integral. 
Since it was assumed that $\hat{\Theta}$ and $\mathbb{P}(\Omega)$ commute, one can work with
$\Gsca(\Omega)$ and $\hat{\Theta}^{\dagger}\hat{\Theta}\Gsca(\Omega)$ instead of evaluating $\hat{\Theta}$ on a vector,
yielding Eq.~\eqref{eq:CPThetabounds}.

The extension to $T > 0$~K follows by noting that $\coth\left( \frac{\hbar\omega}{2k_{B}T} \right)$ has poles at $\omega_{n}$ satisfying
$
    \frac{\hbar\omega_{n}}{2k_{B}T} = i\pi n
$
for $n\in \mathbb{Z}.$ Thus, closing the contour $\int_{-\infty}^{\infty}\mathrm{d}\omega$ into the upper-half plane picks up poles with $\textrm{Im}[\omega_{n}] \geq 0,$ so one only needs to consider $n = 0, 1, \cdots.$
The residue of $\coth\left( \frac{\hbar\omega}{2k_{B}T} \right)$ at the poles as a function $\omega$ is $\frac{2k_{B}T}{\hbar}.$
The indented path around the simple pole at $\omega = 0$ contributes half the residue that a full circle does.
Therefore, if the actual observable or its bound at zero temperature is written as
$F(0) = \int_{0}^{\infty} f(i\xi) \frac{\mathrm{d}\xi}{2\pi}$
for the corresponding integrand $f(i\xi)$, then the corresponding
quantity for temperatures $T > 0$~K is $F(T) = \frac{k_{\mathrm{B}} T}{\hbar} \sum_{n
  = 0}^{'\infty} f(i\xi_{n})$ where the Matsubara frequencies are
$\xi_{n} = 2\pi k_{\mathrm{B}} Tn/\hbar$, and the prime on the
summation means a prefactor of $1/2$ for the contribution at $n=0$.

\begin{figure}
\centering
\includegraphics[width=0.95\columnwidth]{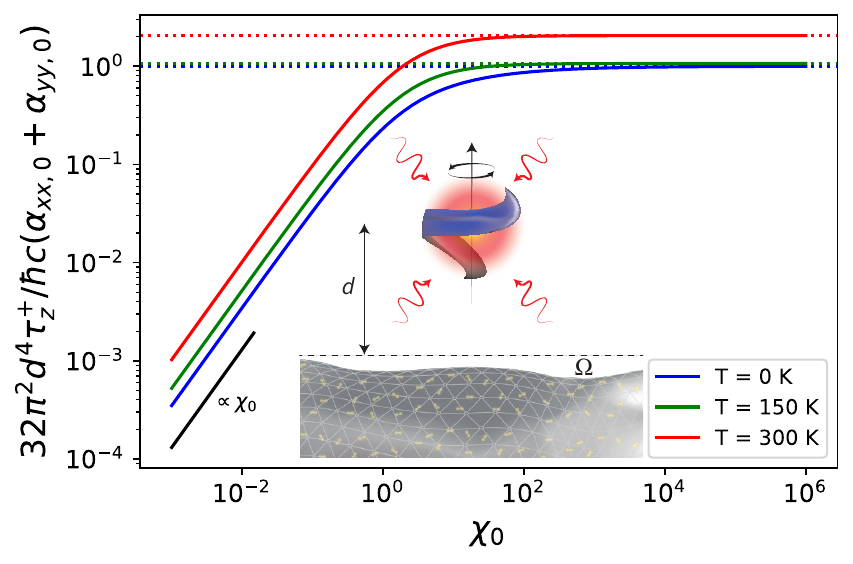}
\caption{
    \textbf{Material dependence of bounds on Casimir torque}.
    Upper bounds (solid lines) on the Casimir torque on a nondispersive dipolar particle above a planar half-space
    design domain $\Omega$ which contains a structure made of nondispersive susceptibility $\chi_{0}$ (inset schematic).
    The bounds saturate in the perfect electric conductor limit (horizontal dotted lines).
    At $T = 0$ K, the bounds approach $\frac{\hbar c}{32\pi^{2} d^{4}}(\alpha_{xx,0} + \alpha_{yy,0})$ as $\chi_{0} \to\infty,$
    and $\frac{7\hbar c}{640\pi^{2} d^{4}}(\alpha_{xx,0} + \alpha_{yy,0})\chi_{0}$ for $\chi_{0} \ll 1$,
    where $d$ is the distance of the particle to the half-space design domain.
    The expression for $T > 0$ K is more complicated, see Eq.~\eqref{eq:tauchi0inflimit}.
    The calculations were done with $d = 10~\mu$m.
    }
\label{fig:torquevschi0}
\end{figure}

\emph{Numerics, asymptotics, and discussion}--- For concreteness, we apply this bound expression on the torque about the $z$ axis exerted by structures contained within a planar semi-infinite half-space  $\Omega = \{(x,y,z) | z \leq 0 \}$ (Fig.~\ref{fig:torquevschi0} inset),
in which case $\Gsca(\Omega)$ and $\hat{J}_{z}^{2}\Gsca(\Omega)$ have known semianalytic forms (App.~\ref{sec:G0Gsca}), allowing for evaluation of
\begin{multline} \label{eq:CPtauzbounds}
  \tau_{\mathrm{dip},z}^{\pm} = \pm \int_{0}^{\infty} \frac{\mathrm{d}\xi}{2\pi} \sum_{a}
  \alpha_{a}
  \sqrt{\frac{\left\langle\vec{u}^{(a)}, \hat{J}_{z}^{2}\Gsca(\Omega)
    \vec{u}^{(a)}\right\rangle }{\left\langle
      \vec{u}^{(a)}, \Gsca(\Omega)
      \vec{u}^{(a)}\right\rangle^{-1}}}.
\end{multline}
This example encompasses a wide variety of possible 
structure designs, but by no means represents the full extent of the theory; bounds can also be evaluated assuming other design domains such as spherical or cylindrical shells. 

For a dipole located at $(0, 0, d)$, we find 
\begin{widetext}
    \begin{multline}
        \tau_{\mathrm{dip},z}^{\pm}(T = 0) = \pm \frac{\hbar c}{16\pi^{2} d^{4}} 
          \int_{0}^{\infty}\int_{0}^{\infty}
          (\alpha_{xx}(\frac{i\tilde{\xi}c}{d}) + \alpha_{yy}(\frac{i\tilde{\xi}c}{d}))\bigg(
          -
          \frac{\tilde{\xi}^{2}}{\sqrt{\tilde{\xi}^{2} + \tilde{k}^{2}}}
          \frac{\sqrt{\tilde{\xi}^{2} + \tilde{k}^{2}} -
            \sqrt{(1 + \chi(\frac{i\tilde{\xi}c}{d}))\tilde{\xi}^{2} + \tilde{k}^{2}}}{\sqrt{\tilde{\xi}^{2} + \tilde{k}^{2}} +
            \sqrt{(1 + \chi(\frac{i\tilde{\xi}c}{d}))\tilde{\xi}^{2} + \tilde{k}^{2}}} \\ 
          +
          \sqrt{\tilde{\xi}^{2} + \tilde{k}^{2}}
          \frac{(1 + \chi(\frac{i\tilde{\xi}c}{d}))\sqrt{\tilde{\xi}^{2} + \tilde{k}^{2}} - \sqrt{(1 +
          \chi(\frac{i\tilde{\xi}c}{d}))\tilde{\xi}^{2} + \tilde{k}^{2}}}{(1 + \chi(\frac{i\tilde{\xi}c}{d}))\sqrt{\tilde{\xi}^{2} + \tilde{k}^{2}}
        + \sqrt{(1 + \chi(\frac{i\tilde{\xi}c}{d}))\tilde{\xi}^{2} + \tilde{k}^{2}}}
          \bigg) 
          e^{-2\sqrt{\tilde{\xi}^{2} + \tilde{k}^{2}}}~\tilde{k}\mathrm{d}\tilde{k}~\mathrm{d}\tilde{\xi}
          \label{eq:torquezexpandedbounds}
    \end{multline}
    \begin{multline}
        \tau_{\mathrm{dip},z}^{\pm}(T > 0) = \pm \frac{k_{B}T}{8\pi d^{3}} 
          \sum_{n=0}^{\infty}{}'
          (\alpha_{xx}(i\xi_{n}) + \alpha_{yy}(i\xi_{n}))
          \int_{0}^{\infty}
          \bigg(
          -
          \frac{(\xi_{n}d/c)^{2}}{\sqrt{(\xi_{n}d/c)^{2} + \tilde{k}^{2}}}
          \frac{\sqrt{(\xi_{n}d/c)^{2} + \tilde{k}^{2}} -
            \sqrt{(1 + \chi(i\xi_{n}))(\xi_{n}d/c)^{2} + \tilde{k}^{2}}}{\sqrt{(\xi_{n}d/c)^{2} + \tilde{k}^{2}} +
            \sqrt{(1 + \chi(i\xi_{n}))(\xi_{n}d/c)^{2} + \tilde{k}^{2}}} \\ 
          +
          \sqrt{(\xi_{n}d/c)^{2} + \tilde{k}^{2}}
          \frac{(1 + \chi(i\xi_{n}))\sqrt{(\xi_{n}d/c)^{2} + \tilde{k}^{2}} - \sqrt{(1 +
          \chi(i\xi_{n}))(\xi_{n}d/c)^{2} + \tilde{k}^{2}}}{(1 + \chi(i\xi_{n}))\sqrt{(\xi_{n}d/c)^{2} + \tilde{k}^{2}}
        + \sqrt{(1 + \chi(i\xi_{n}))(\xi_{n}d/c)^{2} + \tilde{k}^{2}}}
          \bigg) 
          e^{-2\sqrt{(\xi_{n}d/c)^{2} + \tilde{k}^{2}}}~\tilde{k}\mathrm{d}\tilde{k}
          \label{eq:torquezexpandedboundsT}
    \end{multline}
\end{widetext}

Assume for simplicity that the polarizabilities are also dispersionless.
In such a case, the bounds on the torque can be written as $\tau_{\mathrm{dip},z}^{\pm}(T=0) = \pm(\alpha_{xx,0} + \alpha_{yy,0})\frac{\hbar c}{16\pi^{2}d^{4}}g[\chi(i\tilde{\xi} c/d)]$
for some dimensionless functional $g$ which depends only on the macroscopic susceptibility $\chi$ evaluated at a frequency dependent on the separation $d.$
We first consider a macroscopic body of dispersionless susceptibility $\chi(i\xi) = \chi_{0}.$
See Fig.~\ref{fig:torquevschi0} for a numerical evaluation of the bounds.
The bounds scale linearly ($\frac{7\hbar c}{640\pi^{2} d^{4}}(\alpha_{xx,0} + \alpha_{yy,0})\chi_{0}$ for $T=0$~K) as a function of $\chi_{0}$ for $\chi_{0} \ll 1$ and goes to 0 as $\chi_{0} \to 0,$ 
as expected since $\chi_{0} = 0$ implies the design domain is only vacuum in which case the equilibrium torque on the isolated dipole is identically 0.
The bounds increase monotonically with $\chi_{0}.$
Monotonicity can also be expected from Eq.~\eqref{eq:CPtauzbounds} since, expanding $\ket{\vec{E}_{\mathrm{inc}}^{(a)}} \equiv \Gvac\ket{\vec{u}^{(a)}}$ in the basis defined in the previous section,
$
    \sqrt{\left\langle\vec{u}^{(a)}, \hat{\Theta}^{\dagger}\Gsca(\Omega)\hat{\Theta}
    \vec{u}^{(a)}\right\rangle / \left\langle
    \vec{u}^{(a)}, \Gsca(\Omega)
    \vec{u}^{(a)}\right\rangle^{-1}}
$
expands to
$
    \sqrt{
    \bigg(\sum_{\mu}\frac{|e_{\mu}^{(a)}|^{2}|\theta_{\mu}|^{2}}{\chi^{-1}+g_{\mu}}\bigg)/
    \bigg(\sum_{\mu}\frac{|e_{\mu}^{(a)}|^{2}}{\chi^{-1}+g_{\mu}}\bigg)^{-1}}
$
from which it is clear that an increase in $\chi(i\xi)$ can only increase the contribution to the bound at a given $\xi.$
This holds for any $\xi$ and, since $\alpha_{a} > 0,$ therefore for the total integrated ($T = 0$ K) or summed ($T > 0$ K) quantities as well.
However, the bounds do not diverge but rather saturate to a finite value in the perfect electrical conductor (PEC) limit $\chi_{0} \to \infty.$
In the PEC limit, the integrals and summations can be done explicitly and we find
    \begin{align}
    \tau_{\mathrm{dip}}^{\textrm{PEC},\pm} = 
    \begin{cases}
        \pm \frac{\hbar c}{32\pi^{2} d^{4}}(\alpha_{xx,0} + \alpha_{yy,0}), & T = 0, \\
        \pm \frac{\hbar c}{32\pi^{2} d^{4}}
        (\alpha_{xx,0} + \alpha_{yy,0})\times \\
        \frac{a}{2}
        \frac{
        1 + (8a^2 -4a - 1)e^{2a} + (8a^2+4a-1)e^{4a} + e^{6a}
        }
        {2(e^{2a} - 1)^{3}}, 
        & T > 0,
    \end{cases}
    \label{eq:tauchi0inflimit}
    \end{align}
where $a \equiv \frac{2\pi k_{B}Td}{\hbar c}.$ For $T \ll \frac{\hbar c}{2\pi k_{B}d},$ we find
\begin{align}
    \frac{\tau_{\mathrm{dip}}^{\pm}(\chi_{0}\to\infty, T >0)}{\tau_{\mathrm{dip}}^{\pm}(\chi_{0}\to\infty, T =0)} = 
        \bigg(
        1
        -
        \frac{1}{45}\bigg(\frac{2\pi k_{B}Td}{\hbar c}\bigg)^{4}
        + O(T^{6})
        \bigg)
        , 
        \label{eq:tauchi0infTsmalllimit}
    \end{align}
while for $T \gg \frac{\hbar c}{2\pi k_{B}d},$
\begin{align}
    \frac{\tau_{\mathrm{dip}}^{\pm}(\chi_{0}\to\infty, T >0)}{\tau_{\mathrm{dip}}^{\pm}(\chi_{0}\to\infty, T =0)} = 
        \bigg(
        \frac{1}{4}\frac{2\pi k_{B} T d}{\hbar c} + O(T^{2})
        \bigg)
        ,
        \label{eq:tauchi0infTbiglimit}
\end{align}
so nonzero temperature $T$ changes the exact $d^{-4}$ distance scaling of the bounds. 
Although the bounds are applicable at any temperature and any appropriate local, homogeneous, isotropic model of material dispersion, the PEC limit, despite resulting in a looser bound, has the benefit of being analytic resulting in easier extraction of scaling behavior.

\begin{figure}
\centering
\includegraphics[width=0.95\columnwidth]{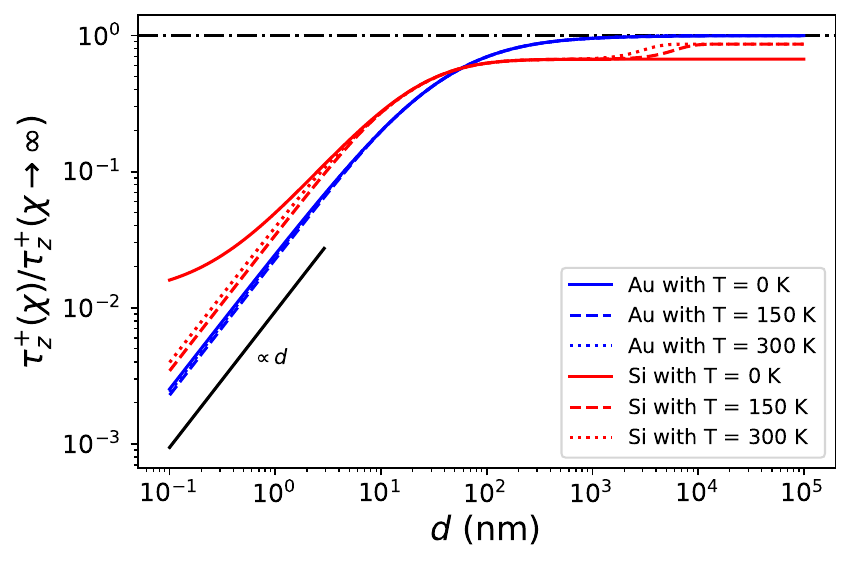}
\caption{\textbf{Distance dependence of bounds on Casimir
    torque for gold and silicon structures}. 
    Upper bounds to the Casimir torque on a nondispersive dipole above a planar half-space design domain which contains a structure with a macroscopic
  susceptibility $\chi$ corresponding to that of gold (Au) or silicon (Si). The results are normalized by the perfect electrical conductor limit given by Eq.~\eqref{eq:tauchi0inflimit}.}
\label{fig:torqueAuvsd}
\end{figure}

Next, we relax the assumption that $\chi(i\xi)$ is nondispersive
and consider the particular cases of i) a gold medium with electric susceptibility modeled by
$\chi_{\mathrm{Au}}(i\xi) = \omega_{\mathrm{p}}^{2} / (\xi^{2} + \gamma\xi)$ where
$\omega_{\mathrm{p}} = 1.37\times 10^{16}~\mathrm{rad/s}$ and $\gamma
= 5.32\times 10^{13}~\mathrm{rad/s}$
and ii) intrinsic (undoped) silicon with $\chi_{\mathrm{Si}}(i\xi) = \epsilon_{\mathrm{Si}}(\infty) - 1 + (\epsilon_{\mathrm{Si}}(0) + \epsilon_{\mathrm{Si}}(\infty))/(1 + \xi^{2}/\omega_{0}^{2})$ where $\epsilon_{\mathrm{Si}}(0) = 11.87$, $\epsilon_{\mathrm{Si}}(\infty) = 1.035$, and $\omega_{0} = 6.6 \times 10^{15}$ rad/s~\cite{venkataram2020fundamentalCPforce}.
For simplicity, we continue to neglect dispersion in $\alpha_{xx}$ and $\alpha_{yy}.$
See Fig.~\ref{fig:torqueAuvsd}. For $d > 1$ $\mu$m, we find that $\tau_{\mathrm{dip},z}^{\pm}(\chi_{\mathrm{Au}})$ is within 5\% of the PEC limit, and within 1\% for $d > 10$ $\mu$m, regardless of the temperature. 

In summary, we presented trace expressions for equilibrium
Casimir torque that apply to arbitrary object shapes and materials,
generalizing prior work on power transfer~\cite{T_operator_bounds_angle_integrated,kruger_nonequilbrium_fluctuations_review,venkataram2020fundamental} and forces~\cite{kruger_nonequilbrium_fluctuations_review,kruger_trace_formulae_for_nonequilibrium,venkataram2020fundamentalCPforce}.
The need for a full account of the spin and orbital angular
momentum carried by waves in this setting is explicit in the trace expressions.
Then, using the derived trace expressions, we calculated bounds on the Casimir torque on a dipolar particle next to a structure of isotropic susceptibility
$\chi$ enclosed within a prescribed domain, and evaluated the
bounds specifically for a half-space design domain as a demonstrative example.
An interesting direction for future work is the extension of the bounds for the case where the structured medium is birefringent or, more generally, anisotropic.

\begin{acknowledgments}

Benjamin Strekha thanks Prashanth S. Venkataram and Pengning Chao for helpful discussions.
This work was supported by the National Science Foundation under the Emerging Frontiers in Research and
Innovation (EFRI) program, EFMA-1640986, the Cornell Center for Materials Research (MRSEC) through award
DMR-1719875, the Defense Advanced Research Projects Agency (DARPA) under agreements HR00112090011,
HR00111820046 and HR0011047197. The views, opinions, and findings expressed herein are those of the 
authors and should not be interpreted as representing the official 
views or policies of any institution.
\end{acknowledgments}

\appendix

\section{Trace expressions for nonreciprocal materials}
\label{app:traceexpressionsderviation}

In this section, we provide the intermediate steps in the derivation of the trace expressions for the torque in terms of the scattering operators and the background Green's function.

\subsection{A single object}

For the case of a single object (reciprocal or nonreciprocal), we use $\mathbb{G} = \mathbb{G}_{0} + \mathbb{G}_{0}\mathbb{T}\mathbb{G}_{0}$ to find that
\begin{align}
    \text{Tr}|_{V}[
    \vecop{J}\mathbb{G}^{\mathsf{A}}\mathbb{G}_{0}^{-1\dagger}
    ]
    &=
    \text{Tr}|_{V}[\vecop{J}(\mathbb{G}_{0}\mathbb{T} - \mathbb{G}_{0}^{\dagger}\mathbb{T}^{\dagger})]\frac{1}{2i}, \\ 
    &=
    \text{Tr}[\vecop{J}(\mathbb{G}_{0}\mathbb{T} - \mathbb{G}_{0}^{\dagger}\mathbb{T}^{\dagger})]\frac{1}{2i}
\end{align}
where the first equality follows since $\mathbb{G}_{0}^{-1} = \nabla\times\nabla\times - \frac{\omega^{2}}{c^{2}}$ inside the body and so gives the identity operator when acting on $\mathbb{G}_{0}$ or $\mathbb{G}_{0}^{\dagger}$ which has one argument restricted to the body. In more detail, $\mathbb{G}_{0}^{-1\dagger} = \mathbb{G}_{0}^{-1} - 2i\mathbb{G}_{0}^{-1\mathsf{A}}$ where $\mathbb{G}_{0}^{-1\mathsf{A}}$ is local and infinitesimal (it is proportional to $\mathbb{V}_{\textrm{env}}^{\mathsf{A}}$).
The $\mathbb{V}_{\textrm{env}}^{\mathsf{A}}$ term is the famous environmental ``dust'' contribution~\cite{kruger_trace_formulae_for_nonequilibrium,eckhardt1984macroscopic}. Its contribution is infinitesimal if integrated over a finite region, but it can be noninfinitesimal if integrated over infinite space.
The second equality follows since the $\mathbb{T}$ operator vanishes unless both arguments are inside the body, so the integration can be extended to the entire space.
Therefore,
\begin{align}
    \boldsymbol{\tau}^{(\textrm{eq})} &= -\textrm{Im}\int_{-\infty}^{\infty}
    \frac{\mathrm{d}\omega}{2\pi}\coth\left(\frac{\hbar\omega}{2k_{B}T} \right)\text{Tr}|_{V}[\hat{\Theta} \mathbb{G}^{\mathsf{A}}\mathbb{G}_{0}^{-1\dagger}], \\
    &= \text{Re}\int_{-\infty}^{\infty} \frac{\mathrm{d}\omega}{4\pi} \coth\left(\frac{\hbar\omega}{2k_{B}T} \right)\text{Tr}[\vecop{J}(\mathbb{G}_{0}\mathbb{T} - \mathbb{G}_{0}^{\dagger}\mathbb{T}^{\dagger})].
\end{align}
Using $\text{Re}\text{Tr}[A] = \text{Re}\text{Tr}[A^{\dagger}]$, $\vecop{J} = \vecop{J}^{\dagger}$, $\vecop{J}\mathbb{G}_{0} = \mathbb{G}_{0}\vecop{J}$, and the cyclicity of the trace,
one finds that $\text{Re}\text{Tr}[\vecop{J}(\mathbb{G}_{0}\mathbb{T} - \mathbb{G}_{0}^{\dagger}\mathbb{T}^{\dagger})] = 0$ for any $\mathbb{T}.$

\begin{widetext}
\subsection{Two or more objects}
We start with
\begin{align}
    \mathbb{G} &= (1 + \mathbb{G}_{0}\mathbb{T}_{\bar{1}})\frac{1}{1 - \mathbb{G}_{0}\mathbb{T}_{1}\mathbb{G}_{0}\mathbb{T}_{\bar{1}}}(1 + \mathbb{G}_{0}\mathbb{T}_{1})\mathbb{G}_{0}, \label{eq:Gexpression1}\\
    &= (1 + \mathbb{G}_{0}\mathbb{T}_{1})\frac{1}{1 - \mathbb{G}_{0}\mathbb{T}_{\bar{1}}\mathbb{G}_{0}\mathbb{T}_{1}}(1 + \mathbb{G}_{0}\mathbb{T}_{\bar{1}})\mathbb{G}_{0}.
    \label{eq:Gexpression2}
\end{align}
The relevant expression for $\mathbb{G}$ is the one for which the right-most $\mathbb{T}$ operator is nonzero in the volume being integrated over. This is because $\mathbb{T}_{\alpha}$ vanishes outside the volume of $\alpha,$ allowing for the extension of the volume integration to be over all space which results in a basis independent trace expression. Collecting terms with $\mathbb{T}_{1}$ as the right-most term, we find
\begin{align}
    \boldsymbol{\tau}^{(1,\textrm{eq})} &= -\textrm{Im}\int_{-\infty}^{\infty}
    \frac{\mathrm{d}\omega}{2\pi}\coth\left(\frac{\hbar\omega}{2k_{B}T} \right)
    \text{Tr}|_{V_{1}}[\vecop{J}\mathbb{G}^{\mathsf{A}}\mathbb{G}_{0}^{-1\dagger}], \\
    &= -\text{Im}\int_{-\infty}^{\infty} \frac{\mathrm{d}\omega}{2\pi} \coth\left(\frac{\hbar\omega}{2k_{B}T} \right)
    \bigg(
    \text{Tr}[\vecop{J}(1 + \mathbb{G}_{0}\mathbb{T}_{\bar{1}})\frac{1}{1 - \mathbb{G}_{0}\mathbb{T}_{1}\mathbb{G}_{0}\mathbb{T}_{\bar{1}}}\mathbb{G}_{0}\mathbb{T}_{1}]
    - \text{Tr}[\vecop{J}\mathbb{G}_{0}^{\dagger}(1 + \mathbb{T}_{\bar{1}}^{\dagger}\mathbb{G}_{0}^{\dagger})\frac{1}{1 - \mathbb{T}_{1}^{\dagger}\mathbb{G}_{0}^{\dagger}\mathbb{T}_{\bar{1}}^{\dagger}\mathbb{G}_{0}^{\dagger}}\mathbb{T}_{1}^{\dagger}]
    \bigg)\frac{1}{2i}, \\
    &= \text{Re}\int_{-\infty}^{\infty} \frac{\mathrm{d}\omega}{4\pi} \coth\left(\frac{\hbar\omega}{2k_{B}T} \right)
    \bigg(
    \text{Tr}[\vecop{J}(1 + \mathbb{G}_{0}\mathbb{T}_{\bar{1}})\frac{1}{1 - \mathbb{G}_{0}\mathbb{T}_{1}\mathbb{G}_{0}\mathbb{T}_{\bar{1}}}\mathbb{G}_{0}\mathbb{T}_{1}]
    - \text{Tr}[\vecop{J}\mathbb{G}_{0}^{\dagger}(1 + \mathbb{T}_{\bar{1}}^{\dagger}\mathbb{G}_{0}^{\dagger})\frac{1}{1 - \mathbb{T}_{1}^{\dagger}\mathbb{G}_{0}^{\dagger}\mathbb{T}_{\bar{1}}^{\dagger}\mathbb{G}_{0}^{\dagger}}\mathbb{T}_{1}^{\dagger}]
    \bigg).
\end{align}
Applying $\text{Re}\text{Tr}[A] = \text{Re}\text{Tr}[A^{\dagger}]$ to the second trace term leads to
\begin{align}
    \boldsymbol{\tau}^{(1,\textrm{eq})} 
    &= \text{Re}\int_{-\infty}^{\infty} \frac{\mathrm{d}\omega}{4\pi} \coth\left(\frac{\hbar\omega}{2k_{B}T} \right)
    \text{Tr}[(\vecop{J}\mathbb{G}_{0}\mathbb{T}_{\bar{1}} - \mathbb{G}_{0}\mathbb{T}_{\bar{1}}\vecop{J})\frac{1}{1 - \mathbb{G}_{0}\mathbb{T}_{1}\mathbb{G}_{0}\mathbb{T}_{2}}\mathbb{G}_{0}\mathbb{T}_{1}], \\
    &=
    \text{Re}\int_{0}^{\infty} \frac{\mathrm{d}\omega}{\pi} [n(\omega, T) + \frac{1}{2}]
    \text{Tr}\bigg[\frac{1}{1 - \mathbb{G}_{0}\mathbb{T}_{1}\mathbb{G}_{0}\mathbb{T}_{\bar{1}}}\mathbb{G}_{0}
    (\mathbb{T}_{1}\vecop{J} - \vecop{J}\mathbb{T}_{1})\mathbb{G}_{0}\mathbb{T}_{\bar{1}}\bigg]
\end{align}
which is precisely Eq.~\eqref{eq:taualphaeqalmostln} when $\alpha = 1.$
It is clear that the derivation for $\bar{1}$ is analogous but with $1 \leftrightarrow \bar{1}$ in intermediate expressions:
\begin{align}
    \boldsymbol{\tau}^{(\bar{1},\textrm{eq})} &=
    -\textrm{Im}\int_{-\infty}^{\infty}
    \frac{\mathrm{d}\omega}{2\pi}\coth\left(\frac{\hbar\omega}{2k_{B}T} \right)
    \text{Tr}|_{V_{\bar{1}}}[\vecop{J}\mathbb{G}^{\mathsf{A}}\mathbb{G}_{0}^{-1\dagger}], \\
    &=
    \text{Re}\int_{0}^{\infty} \frac{\mathrm{d}\omega}{\pi} [n(\omega, T) + \frac{1}{2}]
    \text{Tr}\bigg[\frac{1}{1 - \mathbb{G}_{0}\mathbb{T}_{\bar{1}}\mathbb{G}_{0}\mathbb{T}_{1}}\mathbb{G}_{0}
    (\mathbb{T}_{\bar{1}}\vecop{J} - \vecop{J}\mathbb{T}_{\bar{1}})\mathbb{G}_{0}\mathbb{T}_{1}\bigg]
\end{align}
which proves Eq.~\eqref{eq:taualphaeqalmostln} for $\alpha = \bar{1}.$
Note that the above equations are valid for reciprocal and nonreciprocal objects since the intermediate steps in the derivation did not assume reciprocity of the $\mathbb{T}$ operators.

\section{Green's function expressions in real space}
\label{sec:G0Gsca}
Here, we provide the expressions for the Green's function necessary
to compute bounds above a half-space design region.

$\Gvac$ is defined as the operator that is the inverse of the
Maxwell operator, satisfying $[(\nabla \times \nabla \times) -
  (\omega/c)^{2} \II]\Gvac(\omega, \vec{x}, \vec{x}') = (\omega/c)^{2}
\II\delta^{3} (\vec{x} - \vec{x}')$. In position space and evaluating
at $\omega = i\xi$, this yields the expression $\Gvac(i\xi,
\vec{x}, \vec{x}') = [\nabla \otimes \nabla + (\xi/c)^{2}](e^{-\xi
  |\vec{x} - \vec{x}'|/c} / 4\pi|\vec{x} - \vec{x}'|)$.

The scattering Green's function at $\omega = i\xi$ in the vacuum
region above a uniform planar semi-infinite half-space of
susceptibility $\chi$ is~\cite{novotny2012principles}
\begin{align}
  \Gsca(i\xi, \vec{x}, \vec{x}') = \frac{1}{2}
  \int_{-\infty}^{\infty} \int_{-\infty}^{\infty}
  (\mathbb{M}^{\mathrm{s}}(i\xi, \vec{k}) +
  \mathbb{M}^{\mathrm{p}}(i\xi, \vec{k})) e^{i(k_{x} (x - x') +
    k_{y} (y - y')) - \kappa_{z} (z +
    z')}~\frac{\mathrm{d}k_{x}~\mathrm{d}k_{y}}{(2\pi)^{2}}
\end{align}
defined in terms of the Cartesian tensors
\begin{align}
  \mathbb{M}^{\mathrm{s}}(i\xi, \vec{k}) &=
  -\frac{\xi^{2} r^{\mathrm{s}}(i\xi, k)}{c^{2} \kappa_{z} k^{2}}
  \begin{bmatrix}
    k_{y}^{2} & -k_{x} k_{y} & 0 \\
    -k_{x} k_{y} & k_{x}^{2} & 0 \\
    0 & 0 & 0
  \end{bmatrix}, \\
  \mathbb{M}^{\mathrm{p}}(i\xi, \vec{k}) &=
  \frac{r^{\mathrm{p}}(i\xi, k)}{k^{2}} \begin{bmatrix}
    \kappa_{z} k_{x}^{2} & \kappa_{z} k_{x} k_{y} & -i k^{2} k_{x} \\
    \kappa_{z} k_{x} k_{y} & \kappa_{z} k_{y}^{2} & -i k^{2} k_{y} \\
    i k^{2} k_{x} & i k^{2} k_{y} & k^{4} / \kappa_{z}
  \end{bmatrix},
\end{align}
which are in turn defined in terms of $\vec{k} = k_{x} \vec{e}_{x} +
k_{y} \vec{e}_{y}$, $k = |\vec{k}|$, $\kappa_{z} = \sqrt{(\xi/c)^{2} +
  k^{2}}$, and the Fresnel reflection coefficients
\begin{align}
  r^{\mathrm{s}}(i\xi, k) &= \frac{\sqrt{(\xi/c)^{2} + k^{2}} -
    \sqrt{(1 + \chi)(\xi/c)^{2} + k^{2}}}{\sqrt{(\xi/c)^{2} + k^{2}} +
    \sqrt{(1 + \chi)(\xi/c)^{2} + k^{2}}}, 
    \\ 
    r^{\mathrm{p}}(i\xi, k) &= \frac{(1 + \chi)\sqrt{(\xi/c)^{2} + k^{2}} - \sqrt{(1 +
      \chi)(\xi/c)^{2} + k^{2}}}{(1 + \chi)\sqrt{(\xi/c)^{2} + k^{2}}
    + \sqrt{(1 + \chi)(\xi/c)^{2} + k^{2}}}
\end{align}
at $\omega = i\xi$.

To evaluate $\Gsca(i\xi, \vec{x}, \vec{x}')$ in torque settings, polar coordinates are a more natural choice. That is, reparameterize $(k_{x}, k_{y})$ as $(k, \psi)$ and $(x,y,z)$ as $(\rho, \phi, z)$ so that
\begin{multline}
  \Gsca(i\xi, \vec{x}, \vec{x}') = \frac{1}{2}
  \int_{0}^{2\pi} \int_{0}^{\infty}
  (\mathbb{M}^{\mathrm{s}}(i\xi, \vec{k}) +
  \mathbb{M}^{\mathrm{p}}(i\xi, \vec{k})) e^{i(k\cos\psi (\rho\cos\phi - \rho'\cos{\phi'}) +
    k\sin\psi (\rho\sin\phi - \rho'\sin{\phi'})) - \kappa_{z} (z +
    z')}~\frac{k\mathrm{d}k~\mathrm{d}\psi}{(2\pi)^{2}}.
\end{multline}
We then find (defining $\hat{R}_{z} \equiv \frac{1}{\hbar}\hat{S}_{z}$)
\begin{multline}
  (\hat{J}_{z}\Gsca)(i\xi, \vec{x}, \vec{x}') = \frac{\hbar}{2}
  \int_{0}^{2\pi} \int_{0}^{\infty}
  \bigg(
  -k\cos\psi\rho\sin\phi + k\sin\psi\rho\cos\phi
  + \hat{R}_{z}
  \bigg)
  (\mathbb{M}^{\mathrm{s}}(i\xi, \vec{k}) +
  \mathbb{M}^{\mathrm{p}}(i\xi, \vec{k})) 
  \\
  \times
  e^{i(k\cos\psi (\rho\cos\phi - \rho'\cos{\phi'}) +
    k\sin\psi (\rho\sin\phi - \rho'\sin{\phi'})) - \kappa_{z} (z +
    z')}~\frac{k\mathrm{d}k~\mathrm{d}\psi}{(2\pi)^{2}}
\end{multline}
and
\begin{multline}
  (\hat{J}_{z}^{2}\Gsca)(i\xi, \vec{x}, \vec{x}') = \frac{\hbar^{2}}{2}
  \int_{0}^{2\pi} \int_{0}^{\infty}
  \bigg(
  (-k\cos\psi\rho\sin\phi + k\sin\psi\rho\cos\phi)^{2}
  - i(-k\cos\psi\rho\cos\phi - k\sin\psi\rho\sin\phi) \\
  + 2\hat{R}_{z}(-k\cos\psi\rho\sin\phi + k\sin\psi\rho\cos\phi)
  + \hat{R}_{z}^{2}
  \bigg)
  \\
  \times (\mathbb{M}^{\mathrm{s}}(i\xi, \vec{k}) +
  \mathbb{M}^{\mathrm{p}}(i\xi, \vec{k})) 
  \\
  \times e^{i(k\cos\psi (\rho\cos\phi - \rho'\cos{\phi'}) +
    k\sin\psi (\rho\sin\phi - \rho'\sin{\phi'})) - \kappa_{z} (z +
    z')}~\frac{k\mathrm{d}k~\mathrm{d}\psi}{(2\pi)^{2}}.
\end{multline}
We are interested in the torque about the center of mass of the dipole. For $\mathbf{x} = \mathbf{x}' = (0, 0, d)$ we get
\begin{align}
  (\hat{J}_{z}^{2}\Gsca)(i\xi, d\mathbf{e}_{z}, d\mathbf{e}_{z}) 
  &= 
  \frac{\hbar^{2}}{2}
  \int_{0}^{2\pi} \int_{0}^{\infty}
  \hat{R}_{z}^{2}
  (\mathbb{M}^{\mathrm{s}}(i\xi, \vec{k}) +
  \mathbb{M}^{\mathrm{p}}(i\xi, \vec{k})) 
  e^{-2\kappa_{z}d}~\frac{k\mathrm{d}k~\mathrm{d}\psi}{(2\pi)^{2}}, \\
  &= 
  \frac{\hbar^{2}}{2}
  \int_{0}^{2\pi} \int_{0}^{\infty}
  \begin{bmatrix}
      1 & 0 & 0 \\
      0 & 1 & 0 \\
      0 & 0 & 0
  \end{bmatrix}
  (\mathbb{M}^{\mathrm{s}}(i\xi, \vec{k}) +
  \mathbb{M}^{\mathrm{p}}(i\xi, \vec{k})) 
  e^{-2\kappa_{z}d}~\frac{k\mathrm{d}k~\mathrm{d}\psi}{(2\pi)^{2}}.
\end{align}
It makes intuitive sense that $(\hat{J}_{z}\Gsca)(i\xi, d\mathbf{e}_{z}, d\mathbf{e}_{z}) = (\hat{S}_{z}\Gsca)(i\xi, d\mathbf{e}_{z}, d\mathbf{e}_{z})$ since a point dipole has no geometric structure so any torque on it must come from internal structure (the polarizability matrix). 
These expressions, after changing to dimensionless integration variables, lead to Eq.~\eqref{eq:torquezexpandedbounds} in the main text. 
Explicitly, in the limit that $\chi \to \infty,$ then $r^{\mathrm{s}} \to -1$ and $r^{\mathrm{p}} \to 1,$ leading to the simplification
\begin{align}
  (\Gsca)(i\xi, d\mathbf{e}_{z}, d\mathbf{e}_{z}) 
  &= 
  \frac{1}{2}
  \int_{0}^{2\pi} \int_{0}^{\infty}
  (
  \frac{\xi^{2}}{c^{2} \kappa_{z} k^{2}}
  \begin{bmatrix}
    k^{2}\sin{\psi}^2 & -k^{2}\cos{\psi}\sin{\psi} & 0 \\
    -k^{2}\cos{\psi}\sin{\psi} & k^{2}\cos{\psi}^2 & 0 \\
    0 & 0 & 0
  \end{bmatrix}
  \nonumber\\
  &\qquad\qquad\qquad +
  \frac{1}{k^{2}} \begin{bmatrix}
    \kappa_{z} k^{2}\cos{\psi}^{2} & \kappa_{z} k^{2}\cos{\psi}\sin{\psi} & -i k^{3}\cos{\psi} \\
    \kappa_{z} k^{2}\cos{\psi}\sin{\psi} & \kappa_{z} k^{2}\sin{\psi}^2 & -i k^{3} \sin{\psi} \\
    i k^{3} \cos{\psi} & i k^{3}\sin{\psi} & k^{4} / \kappa_{z}
  \end{bmatrix}
  ) 
  e^{-2\kappa_{z}d}~\frac{k\mathrm{d}k~\mathrm{d}\psi}{(2\pi)^{2}}, \\
  &=
  \frac{1}{32\pi d^{3}}(1 + 2(\frac{\xi d}{c}) + 4(\frac{\xi d}{c})^2)
  \exp(-2\frac{\xi d}{c})
  \begin{bmatrix}
      1 & 0 & 0 \\
      0 & 1 & 0 \\
      0 & 0 & 0
  \end{bmatrix} \nonumber \\
  &+
  \frac{1}{16\pi d^{3}}(1 + 2(\frac{\xi d}{c}))
  \exp(-2\frac{\xi d}{c})
  \begin{bmatrix}
      0 & 0 & 0 \\
      0 & 0 & 0 \\
      0 & 0 & 1
  \end{bmatrix}
\end{align}
In the $\chi \to \infty$ limit, $(\hat{J}_{z}^{2}\Gsca(i\xi, d\mathbf{e}_{z}, d\mathbf{e}_{z}))$ is the same as the upper-left 2-by-2 block as above but with an additional factor of $\hbar^{2}$.
These expressions lead directly to Eq.~\eqref{eq:tauchi0inflimit} in the main text after doing the remaining integration over $\xi.$

\end{widetext}
\bibliography{refs}

\begin{thebibliography}{51}
\providecommand{\natexlab}[1]{#1}
\providecommand{\url}[1]{\texttt{#1}}
\expandafter\ifx\csname urlstyle\endcsname\relax
  \providecommand{\doi}[1]{doi: #1}\else
  \providecommand{\doi}{doi: \begingroup \urlstyle{rm}\Url}\fi

\bibitem[Klimchitskaya et~al.(2009)Klimchitskaya, Mohideen, and
  Mostepanenko]{klimchitskaya2009casimir}
GL~Klimchitskaya, U~Mohideen, and VM~Mostepanenko.
\newblock The {C}asimir force between real materials: Experiment and theory.
\newblock \emph{Reviews of Modern Physics}, 81\penalty0 (4):\penalty0 1827,
  2009.

\bibitem[Woods et~al.(2016)Woods, Dalvit, Tkatchenko, Rodriguez-Lopez,
  Rodriguez, and Podgornik]{woods2016materials}
LM~Woods, Diego Alejandro~Roberto Dalvit, Alexandre Tkatchenko,
  P~Rodriguez-Lopez, Alejandro~W Rodriguez, and R~Podgornik.
\newblock Materials perspective on {C}asimir and van der {W}aals interactions.
\newblock \emph{Reviews of Modern Physics}, 88\penalty0 (4):\penalty0 045003,
  2016.

\bibitem[Rousseau et~al.(2009)Rousseau, Siria, Jourdan, Volz, Comin, Chevrier,
  and Greffet]{rousseau2009radiative}
Emmanuel Rousseau, Alessandro Siria, Guillaume Jourdan, Sebastian Volz, Fabio
  Comin, Jo{\"e}l Chevrier, and Jean-Jacques Greffet.
\newblock Radiative heat transfer at the nanoscale.
\newblock \emph{Nature Photonics}, 3\penalty0 (9):\penalty0 514--517, 2009.

\bibitem[Cuevas and Garc{\'\i}a-Vidal(2018)]{cuevas2018radiative}
Juan~Carlos Cuevas and Francisco~J Garc{\'\i}a-Vidal.
\newblock Radiative heat transfer.
\newblock \emph{ACS Photonics}, 5\penalty0 (10):\penalty0 3896--3915, 2018.

\bibitem[Lamoreaux(1997)]{lamoreaux1997demonstration}
Steven~K Lamoreaux.
\newblock Demonstration of the {C}asimir force in the 0.6 to 6 $\mu$m range.
\newblock \emph{Physical Review Letters}, 78\penalty0 (1):\penalty0 5, 1997.

\bibitem[Mohideen and Roy(1998)]{mohideen1998precision}
Umar Mohideen and Anushree Roy.
\newblock Precision measurement of the {C}asimir force from 0.1 to 0.9 $\mu$m.
\newblock \emph{Physical Review Letters}, 81\penalty0 (21):\penalty0 4549,
  1998.

\bibitem[Munday et~al.(2009)Munday, Capasso, and Parsegian]{munday2009measured}
Jeremy~N Munday, Federico Capasso, and V~Adrian Parsegian.
\newblock Measured long-range repulsive {C}asimir--{L}ifshitz forces.
\newblock \emph{Nature}, 457\penalty0 (7226):\penalty0 170--173, 2009.

\bibitem[Kats(1971)]{kats1971nonisotropic}
E.~I. Kats.
\newblock Van der {W}aals forces in non-isotropic systems.
\newblock \emph{Sov. Phys. JETP}, 33:\penalty0 634, 1971.

\bibitem[Parsegian and Weiss(1972)]{parsegian1972dielectric}
VA~Parsegian and George~H Weiss.
\newblock Dielectric anisotropy and the van der {W}aals interaction between
  bulk media.
\newblock \emph{The Journal of Adhesion}, 3\penalty0 (4):\penalty0 259--267,
  1972.

\bibitem[Khandekar et~al.(2021)Khandekar, Buddhiraju, Wilkinson, Gimzewski,
  Rodriguez, Chase, and Fan]{chinmay2021forcetorque}
Chinmay Khandekar, Siddharth Buddhiraju, Paul~R. Wilkinson, James~K. Gimzewski,
  Alejandro~W. Rodriguez, Charles Chase, and Shanhui Fan.
\newblock Nonequilibrium lateral force and torque by thermally excited
  nonreciprocal surface electromagnetic waves.
\newblock \emph{Phys. Rev. B}, 104:\penalty0 245433, Dec 2021.
\newblock \doi{10.1103/PhysRevB.104.245433}.

\bibitem[Guo and Fan(2021)]{guo2021single}
Yu~Guo and Shanhui Fan.
\newblock Single gyrotropic particle as a heat engine.
\newblock \emph{ACS Photonics}, 8\penalty0 (6):\penalty0 1623--1629, 2021.

\bibitem[Gao et~al.(2021)Gao, Khandekar, Jacob, and Li]{gao2021thermal}
Xingyu Gao, Chinmay Khandekar, Zubin Jacob, and Tongcang Li.
\newblock Thermal equilibrium spin torque: Near-field radiative angular
  momentum transfer in magneto-optical media.
\newblock \emph{Physical Review B}, 103\penalty0 (12):\penalty0 125424, 2021.

\bibitem[Strekha et~al.(2022)Strekha, Molesky, Chao, Kr{\"u}ger, and
  Rodriguez]{strekha2022tracenoneq}
Benjamin Strekha, Sean Molesky, Pengning Chao, Matthias Kr{\"u}ger, and
  Alejandro~W Rodriguez.
\newblock Trace expressions and associated limits for nonequilibrium {C}asimir
  torque.
\newblock \emph{Physical Review A}, 106\penalty0 (4):\penalty0 042222, 2022.

\bibitem[Munday et~al.(2005)Munday, Iannuzzi, Barash, and
  Capasso]{munday2005torque}
Jeremy~N Munday, Davide Iannuzzi, Yuri Barash, and Federico Capasso.
\newblock Torque on birefringent plates induced by quantum fluctuations.
\newblock \emph{Physical Review A}, 71\penalty0 (4):\penalty0 042102, 2005.

\bibitem[Rodrigues et~al.(2006)Rodrigues, Neto, Lambrecht, and
  Reynaud]{rodrigues2006vacuum}
Robson~B Rodrigues, PA~Maia Neto, A~Lambrecht, and S~Reynaud.
\newblock Vacuum-induced torque between corrugated metallic plates.
\newblock \emph{EPL (Europhysics Letters)}, 76\penalty0 (5):\penalty0 822,
  2006.

\bibitem[Gu{\'e}rout et~al.(2015)Gu{\'e}rout, Genet, Lambrecht, and
  Reynaud]{guerout2015casimir}
R~Gu{\'e}rout, C~Genet, A~Lambrecht, and Serge Reynaud.
\newblock {C}asimir torque between nanostructured plates.
\newblock \emph{EPL (Europhysics Letters)}, 111\penalty0 (4):\penalty0 44001,
  2015.

\bibitem[Somers and Munday(2015)]{somers2015rotation}
David~AT Somers and Jeremy~N Munday.
\newblock Rotation of a liquid crystal by the {C}asimir torque.
\newblock \emph{Physical Review A}, 91\penalty0 (3):\penalty0 032520, 2015.

\bibitem[Somers et~al.(2018)Somers, Garrett, Palm, and
  Munday]{somers2018measurement}
David~AT Somers, Joseph~L Garrett, Kevin~J Palm, and Jeremy~N Munday.
\newblock Measurement of the {C}asimir torque.
\newblock \emph{Nature}, 564\penalty0 (7736):\penalty0 386--389, 2018.

\bibitem[Xu and Li(2017)]{xu2017detecting}
Zhujing Xu and Tongcang Li.
\newblock Detecting {C}asimir torque with an optically levitated nanorod.
\newblock \emph{Physical Review A}, 96\penalty0 (3):\penalty0 033843, 2017.

\bibitem[Ahn et~al.(2018)Ahn, Xu, Bang, Deng, Hoang, Han, Ma, and
  Li]{ahn2018optically}
Jonghoon Ahn, Zhujing Xu, Jaehoon Bang, Yu-Hao Deng, Thai~M Hoang, Qinkai Han,
  Ren-Min Ma, and Tongcang Li.
\newblock Optically levitated nanodumbbell torsion balance and ghz
  nanomechanical rotor.
\newblock \emph{Physical Review Letters}, 121\penalty0 (3):\penalty0 033603,
  2018.

\bibitem[Ahn et~al.(2020)Ahn, Xu, Bang, Ju, Gao, and Li]{ahn2020ultrasensitive}
Jonghoon Ahn, Zhujing Xu, Jaehoon Bang, Peng Ju, Xingyu Gao, and Tongcang Li.
\newblock Ultrasensitive torque detection with an optically levitated
  nanorotor.
\newblock \emph{Nature Nanotechnology}, 15\penalty0 (2):\penalty0 89--93, 2020.

\bibitem[Rahi et~al.(2009)Rahi, Emig, Graham, Jaffe, and
  Kardar]{rahi2009scattering}
Sahand~Jamal Rahi, Thorsten Emig, Noah Graham, Robert~L Jaffe, and Mehran
  Kardar.
\newblock Scattering theory approach to electrodynamic {C}asimir forces.
\newblock \emph{Physical Review D}, 80\penalty0 (8):\penalty0 085021, 2009.

\bibitem[Emig et~al.(2007)Emig, Graham, Jaffe, and Kardar]{emig2007casimir}
Thorsten Emig, N~Graham, RL~Jaffe, and M~Kardar.
\newblock {C}asimir forces between arbitrary compact objects.
\newblock \emph{Physical Review Letters}, 99\penalty0 (17):\penalty0 170403,
  2007.

\bibitem[Gelbwaser-Klimovsky et~al.(2022)Gelbwaser-Klimovsky, Graham, Kardar,
  and Kr{\"u}ger]{gelbwaser2022equilibrium}
David Gelbwaser-Klimovsky, Noah Graham, Mehran Kardar, and Matthias Kr{\"u}ger.
\newblock Equilibrium forces on nonreciprocal materials.
\newblock \emph{Physical Review B}, 106\penalty0 (11):\penalty0 115106, 2022.

\bibitem[Rytov et~al.(1989)Rytov, Kravtsov, and Tatarskii]{rytov1989principles}
Sergei~M Rytov, Yurii~A Kravtsov, and Valeryan~I Tatarskii.
\newblock \emph{Principles of Statistical Radiophysics: Elements of random
  fields}.
\newblock Springer, 1989.

\bibitem[Otey et~al.(2014)Otey, Zhu, Sandhu, and Fan]{otey2014fluctuational}
Clayton~R Otey, Linxiao Zhu, Sunil Sandhu, and Shanhui Fan.
\newblock Fluctuational electrodynamics calculations of near-field heat
  transfer in non-planar geometries: A brief overview.
\newblock \emph{Journal of Quantitative Spectroscopy and Radiative Transfer},
  132:\penalty0 3--11, 2014.

\bibitem[Ding et~al.(2022)Ding, Kollipara, Kim, Kotnala, Li, Chen, and
  Zheng]{ding2022universal}
Hongru Ding, Pavana~Siddhartha Kollipara, Youngsun Kim, Abhay Kotnala, Jingang
  Li, Zhihan Chen, and Yuebing Zheng.
\newblock Universal optothermal micro/nanoscale rotors.
\newblock \emph{Science Advances}, 8\penalty0 (24):\penalty0 eabn8498, 2022.

\bibitem[van~der Laan et~al.(2021)van~der Laan, Tebbenjohanns, Reimann,
  Vijayan, Novotny, and Frimmer]{van2021sub}
Fons van~der Laan, Felix Tebbenjohanns, Ren{\'e} Reimann, Jayadev Vijayan,
  Lukas Novotny, and Martin Frimmer.
\newblock Sub-{K}elvin feedback cooling and heating dynamics of an optically
  levitated librator.
\newblock \emph{Physical Review Letters}, 127\penalty0 (12):\penalty0 123605,
  2021.

\bibitem[Stickler et~al.(2021)Stickler, Hornberger, and
  Kim]{stickler2021quantum}
Benjamin~A Stickler, Klaus Hornberger, and MS~Kim.
\newblock Quantum rotations of nanoparticles.
\newblock \emph{Nature Reviews Physics}, 3\penalty0 (8):\penalty0 589--597,
  2021.

\bibitem[Molesky et~al.(2018)Molesky, Lin, Piggott, Jin, Vuckovi{\'c}, and
  Rodriguez]{molesky2018inverse}
Sean Molesky, Zin Lin, Alexander~Y Piggott, Weiliang Jin, Jelena Vuckovi{\'c},
  and Alejandro~W Rodriguez.
\newblock Inverse design in nanophotonics.
\newblock \emph{Nature Photonics}, 12\penalty0 (11):\penalty0 659--670, 2018.

\bibitem[Christiansen and Sigmund(2021)]{christiansen2021inverse}
Rasmus~E Christiansen and Ole Sigmund.
\newblock Inverse design in photonics by topology optimization: tutorial.
\newblock \emph{JOSA B}, 38\penalty0 (2):\penalty0 496--509, 2021.

\bibitem[Molesky et~al.(2019)Molesky, Jin, Venkataram, and
  Rodriguez]{T_operator_bounds_angle_integrated}
Sean Molesky, Weiliang Jin, Prashanth~S. Venkataram, and Alejandro~W.
  Rodriguez.
\newblock $\mathbb{T}$ operator bounds on angle-integrated absorption and
  thermal radiation for arbitrary objects.
\newblock \emph{Phys. Rev. Lett.}, 123:\penalty0 257401, Dec 2019.
\newblock \doi{10.1103/PhysRevLett.123.257401}.

\bibitem[Molesky et~al.(2020{\natexlab{a}})Molesky, Chao, Jin, and
  Rodriguez]{global_T_operator_bounds}
Sean Molesky, Pengning Chao, Weiliang Jin, and Alejandro~W. Rodriguez.
\newblock Global $\mathbb{T}$ operator bounds on electromagnetic scattering:
  Upper bounds on far-field cross sections.
\newblock \emph{Phys. Rev. Research}, 2:\penalty0 033172, Jul
  2020{\natexlab{a}}.
\newblock \doi{10.1103/PhysRevResearch.2.033172}.

\bibitem[Chao et~al.(2022)Chao, Strekha, Kuate~Defo, Molesky, and
  Rodriguez]{chao2022physical}
Pengning Chao, Benjamin Strekha, Rodrick Kuate~Defo, Sean Molesky, and
  Alejandro~W. Rodriguez.
\newblock Physical limits in electromagnetism.
\newblock \emph{Nature Reviews Physics}, pages 1--17, July 2022.
\newblock ISSN 2522-5820.
\newblock \doi{10.1038/s42254-022-00468-w}.
\newblock Publisher: Nature Publishing Group.

\bibitem[Venkataram et~al.(2020{\natexlab{a}})Venkataram, Molesky, Chao, and
  Rodriguez]{venkataram2020fundamentalCPforce}
Prashanth~S Venkataram, Sean Molesky, Pengning Chao, and Alejandro~W Rodriguez.
\newblock Fundamental limits to attractive and repulsive {C}asimir-{P}older
  forces.
\newblock \emph{Physical Review A}, 101\penalty0 (5):\penalty0 052115,
  2020{\natexlab{a}}.

\bibitem[Strekha et~al.(2023)Strekha, Chao, Defo, Molesky, and
  Rodriguez]{strekhachao2023minldos}
Benjamin Strekha, Pengning Chao, Rodrick~Kuate Defo, Sean Molesky, and
  Alejandro~W Rodriguez.
\newblock Suppressing electromagnetic local density of states via slow light in
  lossy quasi-1d gratings.
\newblock \emph{arXiv preprint arXiv:2309.15794}, 2023.

\bibitem[Kr\"uger et~al.(2012)Kr\"uger, Bimonte, Emig, and
  Kardar]{kruger_trace_formulae_for_nonequilibrium}
Matthias Kr\"uger, Giuseppe Bimonte, Thorsten Emig, and Mehran Kardar.
\newblock Trace formulas for nonequilibrium {C}asimir interactions, heat
  radiation, and heat transfer for arbitrary objects.
\newblock \emph{Phys. Rev. B}, 86:\penalty0 115423, Sep 2012.
\newblock \doi{10.1103/PhysRevB.86.115423}.

\bibitem[Zhang et~al.(2022)Zhang, Zhu, Zhang, and Wang]{zhang2022microscopic}
Yong-Mei Zhang, Tao Zhu, Zu-Quan Zhang, and Jian-Sheng Wang.
\newblock Microscopic theory of photon-induced energy, momentum, and angular
  momentum transport in the nonequilibrium regime.
\newblock \emph{Physical Review B}, 105\penalty0 (20):\penalty0 205421, 2022.

\bibitem[Wang et~al.(2023)Wang, Peng, Zhang, Zhang, and Zhu]{wang2023transport}
Jian-Sheng Wang, Jiebin Peng, Zu-Quan Zhang, Yong-Mei Zhang, and Tao Zhu.
\newblock Transport in electron-photon systems.
\newblock \emph{Frontiers of Physics}, 18\penalty0 (4):\penalty0 43602, 2023.

\bibitem[Wang and Antezza(2023)]{wang2023photon}
Jian-Sheng Wang and Mauro Antezza.
\newblock Photon mediated energy, linear and angular momentum transport in
  fullerene and graphene systems beyond local equilibrium.
\newblock \emph{arXiv preprint arXiv:2307.11361}, 2023.

\bibitem[Khersonskii et~al.(1989)Khersonskii, Moskalev, and
  Varshalovich]{khersonskii1988quantum}
V.K. Khersonskii, A.N. Moskalev, and D.A. Varshalovich.
\newblock \emph{Quantum Theory Of Angular Momemtum}.
\newblock World Scientific Publishing Company, 1989.
\newblock ISBN 9789814578288.

\bibitem[Bimonte et~al.(2017)Bimonte, Emig, Kardar, and
  Krüger]{kruger_nonequilbrium_fluctuations_review}
Giuseppe Bimonte, Thorsten Emig, Mehran Kardar, and Matthias Krüger.
\newblock Nonequilibrium fluctuational quantum electrodynamics: Heat radiation,
  heat transfer, and force.
\newblock \emph{Annual Review of Condensed Matter Physics}, 8\penalty0
  (1):\penalty0 119--143, 2017.
\newblock \doi{10.1146/annurev-conmatphys-031016-025203}.

\bibitem[Novotny and Hecht(2012)]{novotny2012principles}
Lukas Novotny and Bert Hecht.
\newblock \emph{Principles of nano-optics}.
\newblock Cambridge university press, 2012.

\bibitem[Boyd and Vandenberghe(2004)]{boyd2004convex}
Stephen Boyd and Lieven Vandenberghe.
\newblock \emph{Convex optimization}.
\newblock Cambridge University Press, 2004.

\bibitem[Jackson(1999)]{jackson1999classical}
John~David Jackson.
\newblock Classical electrodynamics, 1999.

\bibitem[Molesky et~al.(2020{\natexlab{b}})Molesky, Chao, and
  Rodriguez]{molesky2020hierarchical}
Sean Molesky, Pengning Chao, and Alejandro~W. Rodriguez.
\newblock Hierarchical mean-field $\mathbb{T}$ operator bounds on
  electromagnetic scattering: Upper bounds on near-field radiative purcell
  enhancement.
\newblock \emph{Physical Review Research}, 2:\penalty0 043398, Dec
  2020{\natexlab{b}}.
\newblock \doi{10.1103/PhysRevResearch.2.043398}.

\bibitem[Amaolo et~al.(2023)Amaolo, Chao, Maldonado, Molesky, and
  Rodriguez]{amaolo2023canheterostructures}
Alessio Amaolo, Pengning Chao, Thomas~J Maldonado, Sean Molesky, and
  Alejandro~W Rodriguez.
\newblock Can photonic heterostructures provably outperform single-material
  geometries?
\newblock \emph{arXiv preprint arXiv:2307.00629}, 2023.

\bibitem[Angeris et~al.(2019)Angeris, Vu{\v{c}}kovi{\'c}, and
  Boyd]{angeris2019computational}
Guillermo Angeris, Jelena Vu{\v{c}}kovi{\'c}, and Stephen~P Boyd.
\newblock Computational bounds for photonic design.
\newblock \emph{ACS Photonics}, 6\penalty0 (5):\penalty0 1232, 2019.
\newblock \doi{10.1021/acsphotonics.9b00154}.

\bibitem[Shim et~al.(2019)Shim, Fan, Johnson, and Miller]{shim2019fundamental}
Hyungki Shim, Lingling Fan, Steven~G Johnson, and Owen~D Miller.
\newblock Fundamental limits to near-field optical response over any bandwidth.
\newblock \emph{Physical Review X}, 9\penalty0 (1):\penalty0 011043, 2019.

\bibitem[Venkataram et~al.(2020{\natexlab{b}})Venkataram, Molesky, Jin, and
  Rodriguez]{venkataram2020fundamental}
Prashanth~S Venkataram, Sean Molesky, Weiliang Jin, and Alejandro~W Rodriguez.
\newblock Fundamental limits to radiative heat transfer: the limited role of
  nanostructuring in the near-field.
\newblock \emph{Physical Review Letters}, 124\penalty0 (1):\penalty0 013904,
  2020{\natexlab{b}}.

\bibitem[Eckhardt(1984)]{eckhardt1984macroscopic}
W~Eckhardt.
\newblock Macroscopic theory of electromagnetic fluctuations and stationary
  radiative heat transfer.
\newblock \emph{Physical Review A}, 29\penalty0 (4):\penalty0 1991, 1984.

\end{thebibliography}


\begin{thebibliography}{7}
\providecommand{\natexlab}[1]{#1}
\providecommand{\url}[1]{\texttt{#1}}
\expandafter\ifx\csname urlstyle\endcsname\relax
  \providecommand{\doi}[1]{doi: #1}\else
  \providecommand{\doi}{doi: \begingroup \urlstyle{rm}\Url}\fi

\bibitem[Asheichyk et~al.(2017)Asheichyk, M\"uller, and
  Kr\"uger]{asheichyk2017RHTPPs}
Kiryl Asheichyk, Boris M\"uller, and Matthias Kr\"uger.
\newblock Heat radiation and transfer for point particles in arbitrary
  geometries.
\newblock \emph{Phys. Rev. B}, 96:\penalty0 155402, Oct 2017.
\newblock \doi{10.1103/PhysRevB.96.155402}.

\bibitem[Ott et~al.(2019)Ott, Messina, Ben-Abdallah, and
  Biehs]{ott2019magnetothermoplasmonics}
Annika Ott, Riccardo Messina, Philippe Ben-Abdallah, and Svend-Age Biehs.
\newblock Magnetothermoplasmonics: from theory to applications.
\newblock \emph{Journal of Photonics for Energy}, 9\penalty0 (3):\penalty0
  032711, 2019.

\bibitem[Golyk et~al.(2013)Golyk, Kr{\"u}ger, and Kardar]{golyk2013linear}
Vladyslav~A Golyk, Matthias Kr{\"u}ger, and Mehran Kardar.
\newblock Linear response relations in fluctuational electrodynamics.
\newblock \emph{Physical Review B}, 88\penalty0 (15):\penalty0 155117, 2013.

\bibitem[Strekha et~al.(2022)Strekha, Molesky, Chao, Kr{\"u}ger, and
  Rodriguez]{strekha2022tracenoneq}
Benjamin Strekha, Sean Molesky, Pengning Chao, Matthias Kr{\"u}ger, and
  Alejandro~W Rodriguez.
\newblock Trace expressions and associated limits for nonequilibrium {C}asimir
  torque.
\newblock \emph{Physical Review A}, 106\penalty0 (4):\penalty0 042222, 2022.

\bibitem[Novotny and Hecht(2012)]{novotny2012principles}
Lukas Novotny and Bert Hecht.
\newblock \emph{Principles of nano-optics}.
\newblock Cambridge university press, 2012.

\bibitem[Bimonte et~al.(2017)Bimonte, Emig, Kardar, and
  Krüger]{kruger_nonequilbrium_fluctuations_review}
Giuseppe Bimonte, Thorsten Emig, Mehran Kardar, and Matthias Krüger.
\newblock Nonequilibrium fluctuational quantum electrodynamics: Heat radiation,
  heat transfer, and force.
\newblock \emph{Annual Review of Condensed Matter Physics}, 8\penalty0
  (1):\penalty0 119--143, 2017.
\newblock \doi{10.1146/annurev-conmatphys-031016-025203}.

\bibitem[Kr\"uger et~al.(2012)Kr\"uger, Bimonte, Emig, and
  Kardar]{kruger_trace_formulae_for_nonequilibrium}
Matthias Kr\"uger, Giuseppe Bimonte, Thorsten Emig, and Mehran Kardar.
\newblock Trace formulas for nonequilibrium {C}asimir interactions, heat
  radiation, and heat transfer for arbitrary objects.
\newblock \emph{Phys. Rev. B}, 86:\penalty0 115423, Sep 2012.
\newblock \doi{10.1103/PhysRevB.86.115423}.

\end{thebibliography}
\end{document}


\title{Trace expressions and associated limits for equilibrium Casimir torque: supplemental material}

\author{Benjamin Strekha}
\affiliation{Department of Electrical and Computer Engineering, Princeton University, Princeton, New Jersey 08544, USA}

\author{Matthias Kr\"{u}ger}
\affiliation{Institute for Theoretical Physics, Georg-August-Universit\"{a}t G\"{o}ttingen, 37073 G\"{o}ttingen, Germany}

\author{Alejandro W. Rodriguez}
\affiliation{Department of Electrical and Computer Engineering, Princeton University, Princeton, New Jersey 08544, USA}

\begin{abstract}
    This document provides supplemental material to ``Trace expressions and associated limits for equilibrium Casimir torque'' which are minor extensions or remarks of the ideas presented in the main text.
\end{abstract}

\maketitle

\section{Torque expression in the point-particle limit in the two-body case}

Using the point particle limit and the Born approximation, a simplified expression for the torque exerted on particle 1 is (assuming reciprocity of the materials)
\begin{align}
    \boldsymbol{\tau}^{(1,\textrm{eq})}(T) &= \text{Re}\frac{2}{\pi} \int_{0}^{\infty} \mathrm{d}\omega [n(\omega, T) + \frac{1}{2}]
    \text{Tr}[
    \vecop{J}\mathbb{G}_{0}\mathbb{T}_{2}
    \mathbb{G}_{0}\mathbb{T}_{1}
    ].
\end{align}
Using the scattering operators in the small-sphere limit~\cite{asheichyk2017RHTPPs}, 
\begin{align}
    \text{Tr}
    [
    \hat{J}_{z}\mathbb{G}_{0}\mathbb{T}_{2}\mathbb{G}_{0}\mathbb{T}_{1}
    ]
    =
    &\text{Re}\frac{9\omega^{4}}{c^{4}}\int_{V_{1}} \mathrm{d}^{3}\mathbf{r}\int_{V_{2}} \mathrm{d}^{3}\mathbf{r'}
    (\hat{J}_{z}\mathbb{G}_{0})_{ab}(\mathbf{r}, \mathbf{r}')
    \left(\frac{\MatrixVariable{\epsilon}_{2} - 1}{\MatrixVariable{\epsilon}_{2}+2}\right)_{bc}
    \mathbb{G}_{0,cd}(\mathbf{r}', \mathbf{r})\left(\frac{\MatrixVariable{\epsilon}_{1} - 1}{\MatrixVariable{\epsilon}_{1}+2}\right)_{da}.
\end{align}
Since the dimensions of the point particles are assumed small compared to any other dimensions in the problem, $(\hat{J}_{z}\mathbb{G}_{0})(\mathbf{r}, \mathbf{r}')$ and $\mathbb{G}_{0}(\mathbf{r}', \mathbf{r})$ do not vary significantly between different points in the different particles. Letting $\mathbf{r}_{1}$ and $\mathbf{r}_{2}$ denote the centers of particle 1 and particle 2, respectively, the integrals $\int_{V_{1}}$ and $\int_{V_{2}}$ simply introduce factors of $4\pi R_{1}^{3}/3$ and $4\pi R_{2}^{3}/3$ so that the torque is proportional to the volumes of the particles. Compactly,
\begin{align}
    \boldsymbol{\tau}^{(1,\textrm{eq})}(T) &= \text{Re}\frac{2}{\pi} \int_{0}^{\infty} \mathrm{d}\omega \frac{\omega^{4}}{c^{4}} [n(\omega, T) + \frac{1}{2}] 
    \text{Tr}_{cmp}[
    \vecop{J}\mathbb{G}_{0}(\mathbf{r}_{1}, \mathbf{r}_{2})
    \MatrixVariable{\alpha}_{2}
    \mathbb{G}_{0}(\mathbf{r}_{2}, \mathbf{r}_{1})
    \MatrixVariable{\alpha}_{1}
    ]
\end{align}
where $\text{Tr}_{cmp}$ means a trace only over the vector components ($\text{Tr}_{cmp}[\mathbb{A}(\mathbf{r}, \mathbf{r}')] \equiv \sum_{a} \mathbb{A}_{aa}(\mathbf{r}, \mathbf{r}')$). For the general nonreciprocal case, similar arguments lead to
\begin{align}
    \boldsymbol{\tau}^{(1,\textrm{eq})}(T) &=
    \text{Re}\frac{1}{\pi}\int_{0}^{\infty} \mathrm{d}\omega [n(\omega, T) + \frac{1}{2}] \text{Tr}\bigg[\mathbb{G}_{0}
    (\mathbb{T}_{1}\vecop{J} - \vecop{J}\mathbb{T}_{1})\mathbb{G}_{0}\mathbb{T}_{2}\bigg], 
    \label{supp:eqtaualphaeqalmostln}
    \\
    &=
    \text{Re}\frac{1}{\pi}\int_{0}^{\infty} \mathrm{d}\omega \frac{\omega^{4}}{c^{4}} [n(\omega, T) + \frac{1}{2}] 
    \text{Tr}_{cmp}\bigg[
    \vecop{J}\mathbb{G}_{0}(\mathbf{r}_{1}, \mathbf{r}_{2})
    \MatrixVariable{\alpha}_{2}
    \mathbb{G}_{0}(\mathbf{r}_{2}, \mathbf{r}_{1})
    \MatrixVariable{\alpha}_{1}
    -
    \vecop{J}\mathbb{G}_{0}(\mathbf{r}_{2}, \mathbf{r}_{1})
    \MatrixVariable{\alpha}_{1}
    \mathbb{G}_{0}(\mathbf{r}_{1}, \mathbf{r}_{2})
    \MatrixVariable{\alpha}_{2}\bigg].
    \label{supp:eqTorqueTwoDipoles}
\end{align}

\section{Numerical examples in the dilute limit}

A natural question is how the magnitude of contributions of the orbital and spin term in the torque expression compare.
The comparison of the contributions of the orbital and spin terms simplifies greatly in the dilute limit, for which one can approximate each $\mathbb{T}$ operator as $\mathbb{T} = \mathbb{V}(1 - \mathbb{G}_{0}\mathbb{V})^{-1} \approx \mathbb{V}.$ In such limits, one can easily create scenarios for which a) $\tau_{L_{z}} \neq 0$ and $\tau_{S_{z}} \neq 0$ b) $\tau_{L_{z}} \neq 0$ but $\tau_{S_{z}} = 0$ c) $\tau_{L_{z}} = 0$ but $\tau_{S_{z}} \neq 0.$

\begin{figure*}
     \centering
     \begin{subfigure}[b]{0.48\textwidth}
         \centering
         \includegraphics[width=\textwidth]{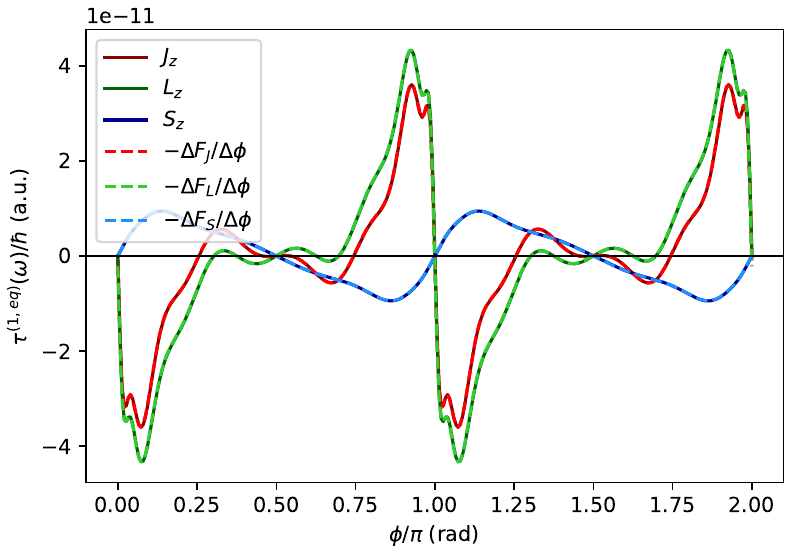}
         \caption{Two birefringent wires.}
         \label{fig:twoWiresTorque_br}
     \end{subfigure}
     \hfill
     \begin{subfigure}[b]{0.48\textwidth}
         \centering
         \includegraphics[width=\textwidth]{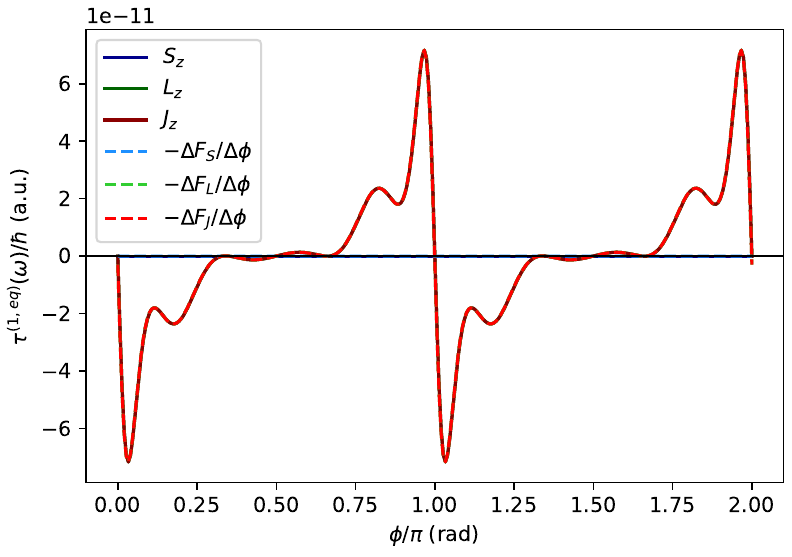}
         \caption{InSb thin wires in the $xy$-plane subject to an external $1.0$ T magnetic field in the $z$-direction.}
         \label{fig:twoWiresTorque_nr}
     \end{subfigure}
     \hfill
     \caption{\textbf{Torque on the top wire as a function of angle $\phi$ between the two wires separated by a fixed distance.} Here, $T = 300$ K, $\omega = k_{B}T/\hbar = 3.92\cdot 10^{13}$, the length of each wire is $L = 4\pi c/\omega = 96.087~\mu$m, and the separation is $d = L/10 = 9.6087~\mu$m.}
     \label{fig:twoWiresTorque}
\end{figure*}

We consider two thin wires parallel to the $x-y$ plane, separated by a distance $d$ in the $z$ direction with an angle of orientation
$\phi$ between the wires ($\phi = 0$ corresponding to parallel wires).
To simplify the numerics, we focus on a single angular frequency $\omega.$ In such a case, one can approximate each $\mathbb{T}$ operator as
\begin{align}
    \mathbb{T}(\mathbf{r}, \mathbf{r}') =
    \begin{cases}
        k^{2}(\MatrixVariable{\epsilon} - \mathbb{I})\delta(\mathbf{r}, \mathbf{r}') & \mathbf{r},\mathbf{r}' \in V_{\textrm{wire}}, \\
        0& \text{else}. \\
    \end{cases}
\end{align}
where $\MatrixVariable{\epsilon}$ is an electric permittivity tensor.
If the permittivity tensor is anisotropic, then both the orbital and spin terms are, in general, nonzero. If the permittivity tensor is isotropic, the spin contribution is exactly 0 for all angles $\phi$ at each frequency $\omega$ while the orbital terms are, in general, nonzero. 
Figure~\ref{fig:twoWiresTorque} shows the torque on the top wire as a function of the orientation
$\phi$ between the wires as well as the magnitude and sign of the contributions of the $\hat{L}_{z}$ and
$\hat{S}_{z}$ terms.
Using finite-difference methods, we calculate $-\Delta \mathcal{F}_{J,L,S}/\Delta\phi$ where $\Delta \mathcal{F}_{J,L,S}$ is the change in free energy due to a slight geometric structure and dielectric rotation of the top wire ($J$), or solely geometric structure rotation ($L$), or solely from the dielectric matrix rotation ($S$), respectively.
We also directly compute the torque using Eq.~\eqref{supp:eqtaualphaeqalmostln} and plot the individual $\hat{L}_{z}$ and $\hat{S}_{z}$ contributions.
The key takeaway is that the $\hat{S}$ contributions can be of the same order of magnitude as the $\hat{L}$ terms, and can have the same or opposite sign.

\begin{figure}
    \centering
    \includegraphics[width=0.45\textwidth]{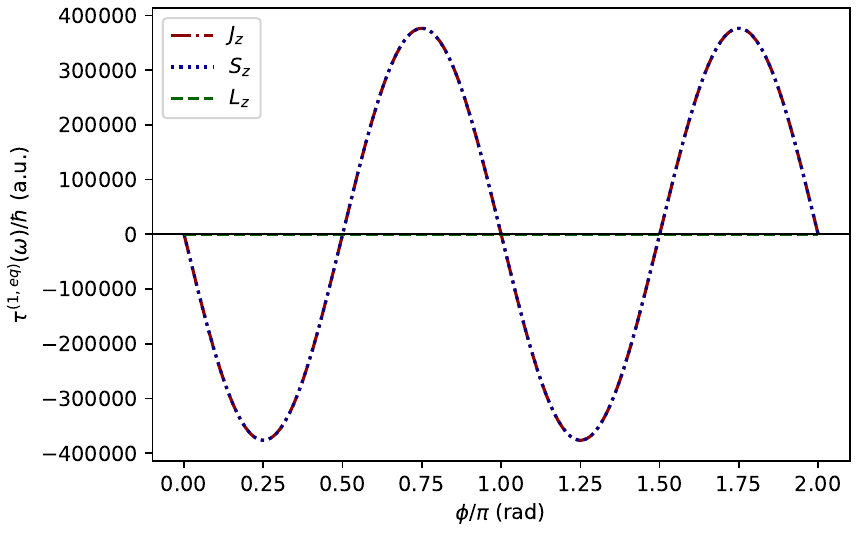} 
    \caption{\textbf{Torque on the top ball as a function of angle $\phi$ of rotation of the top ball.} Here, $T = 300$ K, $R = 100$ nm, $L = 4\pi c/\omega = 48.04~\mu$m, and the separation is $d = 1~\mu$m.}
    \label{fig:twoBallsTorque}
\end{figure}

For the case c) for which $\tau_{L_{z}} = 0$ but $\tau_{S_{z}} \neq 0$ perhaps the simplest example is that of two sub-thermal-wavelength balls (particles) which can be described by polarizability tensors $\MatrixVariable{\alpha}_{1}$ and $\MatrixVariable{\alpha}_{2}(\omega) \equiv 4\pi
R_{2}^{3}\frac{\MatrixVariable{\epsilon}_{2}(\omega) -
  \mathbb{I}}{\MatrixVariable{\epsilon}_{2}(\omega) + 2\mathbb{I}}$ located a positions $\mathbf{r}_{1}$ and $\mathbf{r}_{2},$
respectively. In such a case, a derived above the torque expression reduces to Eq.~\eqref{supp:eqTorqueTwoDipoles}. 
We consider particles with permittivities
\begin{align}
    \MatrixVariable{\epsilon} = 
    \begin{bmatrix}
    \epsilon_{1} & -i\epsilon_{2} & 0 \\
    i\epsilon_{2} & \epsilon_{1} & 0 \\
    0 & 0 & \epsilon_{3}
    \end{bmatrix}_{cart}.
\end{align}
For InSb one has~\cite{ott2019magnetothermoplasmonics}
\begin{align}
    \epsilon_{1} &= \epsilon_{\infty}
    \bigg(
    1 + \frac{\omega_{L}^{2} - \omega_{T}^{2}}{\omega_{T}^{2} - \omega^{2} - i\Gamma\omega} +
    \frac{\omega_{p}^{2}(\omega + i\gamma)}{\omega[\omega_{c}^{2} - (\omega + i\gamma)^{2}]}
    \bigg), \\
    \epsilon_{2} &= \frac{\epsilon_{\infty}\omega_{p}^{2}\omega_{c}}{\omega[(\omega + i\gamma)^{2} - \omega_{c}^{2}]}, \\
    \epsilon_{3} &=
    \epsilon_{\infty}
    \bigg(
    1 + \frac{\omega_{L}^{2} - \omega_{T}^{2}}{\omega_{T}^{2} - \omega^{2} - i\Gamma\omega} -
    \frac{\omega_{p}^{2}}{\omega(\omega + i\gamma)}
    \bigg),
\end{align}
where $\epsilon_{\infty} = 15.7, \omega_{L} = 3.62\times 10^{13}$ rad/s, $\omega_{T} = 3.39\times 10^{13}$ rad/s, $n = 1.07\times 10^{17}$ cm$^{-3}, m^{*} = 1.99\times 10^{-32}$ kg, $\omega_{p} = \sqrt{\frac{nq^{2}}{m^{*}\epsilon_{0}\epsilon_{\infty}}} = 3.15\times 10^{13}$ rad/s, $q = 1.6\times 10^{-19} $ C, $\Gamma = 5.65 \times 10^{11}$ rad/s, $\gamma = 3.39\times 10^{12}$ rad/s, and $\omega_{c} = \frac{eB}{m^{*}}.$
See Fig.~\ref{fig:twoBallsTorque}.

\section{Linear response and fluctuational electrodynamics}

Near equilibrium, linear response functions can be related to fluctuations of corresponding equilibrium quantities.
This was verified to hold in the $\mathbb{T}$ operator trace formalism for fluctuational electrodynamics for heat transfer and forces in Ref.~\cite{golyk2013linear}. Using the newly derived trace expressions for equilibrium Casimir torque along with the nonequilibrium Casimir torque trace expressions from Ref.~\cite{strekha2022tracenoneq} we verified that, unsurprisingly, similar expressions hold for torque.

Let $H^{(\beta)}(t)$ denote the absorbed power by object $\beta.$ It can be written as an integral over the volume $V_{\beta}$ of the local power absorption, which is equal to the inner product of the electric field $\mathbf{E}(\mathbf{r}, t)$ and the current density $\mathbf{K}(\mathbf{r}, t)$
\begin{align}
    H^{(\beta)}(t) = \int_{V_{\beta}} \mathrm{d}^{3}\mathbf{r} ~ 
    \{E_{a}(\mathbf{r}, t), K_{a}(\mathbf{r}, t)\}_{S}
    \label{eq:Hvolumeintdef}
\end{align}
where we use the Einstein summation convention.
Here, $\{A, B\}_{S} \equiv (AB + BA)/2$ is the symmetrized product of two, in general, noncommuting operators.

A general operator $\hat{\Theta}^{(\beta)}(t)$ can be written as
\begin{align}
    \hat{\Theta}^{(\beta)}(t) \equiv &\int_{\mathbf{r}\in V_{\beta}} \mathrm{d}^{3}\mathbf{r}\int_{-\infty}^{\infty} \mathrm{d}\omega\int_{-\infty}^{\infty} \mathrm{d}\omega' 
    e^{-i(\omega + \omega')t}\frac{1}{\hbar\omega}
    \{K_{a}(\mathbf{r}, \omega'), \hat{\Theta} E_{a}(\mathbf{r}, \omega)\}_{S}.
\end{align}
We seek to evaluate
\begin{align}
    \int_{0}^{\infty} \mathrm{d}t \langle\hat{\Theta}^{(\beta)}(t)H^{(\alpha)}(0)\rangle_{\mathrm{eq}}
    \label{eq:dtThetaalphaHbeta}
\end{align}
where $H^{\alpha}$ is as defined in Eq.~\eqref{eq:Hvolumeintdef}.
Using Wick's theorem and simplifying, we find that Eq.~\eqref{eq:dtThetaalphaHbeta} is equal to
\begin{align}
    \int_{0}^{\infty} \mathrm{d}\omega \int_{\mathbf{r}\in V_{\beta}}\int_{\mathbf{r}' \in V_{\alpha}}&\frac{2\pi}{\hbar\omega}
    [
    \langle K_{b}(\mathbf{r})K_{a}^{*}(\mathbf{r}')\rangle_{-\omega} \langle\hat{\Theta}E_{b}(\mathbf{r})E_{a}^{*}(\mathbf{r}')\rangle_{\omega}
    +
    \langle K_{b}(\mathbf{r})E_{a}^{*}(\mathbf{r}')\rangle_{-\omega} \langle\hat{\Theta}E_{b}(\mathbf{r})K_{a}^{*}(\mathbf{r}')\rangle_{\omega}
    ].
\end{align}
Note that $\int_{0}^{\infty} \mathrm{d}t \langle \{A(t), B(0)\}_{S}\rangle_{\mathrm{eq}} = \int_{0}^{\infty} \mathrm{d}t \langle A(t)B(0)\rangle_{\mathrm{eq}}$, so we now work with unsymmetrized correlators to reduce the number of terms from Wick's theorem by half.
Using $\mathbf{E} = i\omega\mu_{0}\mathbb{G}_{0}\mathbf{K}$ along with the unsymmetrical correlator~\cite{novotny2012principles}
\begin{align}
    \langle E_{i}(\mathbf{r},
\omega)E_{j}^{*}(\mathbf{r}', \omega ) \rangle_{\omega} =
\frac{\hbar\omega^{2}}{\pi c^{2}\epsilon_{0}} \frac{1}{1 - e^{-\frac{\hbar\omega}{k_{B}T}}}
\mathbb{G}^{\mathsf{A}}_{ij}(\omega; \mathbf{r}, \mathbf{r}')
\end{align}
we find
\begin{align}
    \int_{0}^{\infty} \mathrm{d}t \langle\hat{\Theta}^{(\beta)}(t)H^{(\alpha)}(0)\rangle_{\mathrm{eq}} &=
    \frac{2}{\pi}\int_{0}^{\infty} \mathrm{d}\omega \frac{\hbar\omega e^{\frac{\hbar\omega}{k_{B}T}}}{(e^{\frac{\hbar\omega}{k_{B}T}} - 1)^{2}}
    \nonumber \\
    &\times[
    \hat{\Theta}\mathbb{G}^{\mathsf{A}}_{ba}(\mathbf{r}, \mathbf{r}')\mathbb{G}_{0,ad}^{-1}(\mathbf{r}', \mathbf{v}) \mathbb{G}^{\mathsf{A}}_{dc}(\mathbf{v}, \mathbf{u})\mathbb{G}_{0,cb}^{-1\dagger}(\mathbf{u}, \mathbf{r})
    \nonumber \\
    &-
    \hat{\Theta}\mathbb{G}^{\mathsf{A}}_{bd}(\mathbf{r}, \mathbf{v})\mathbb{G}_{0,da}^{-1\dagger}(\mathbf{v}, \mathbf{r}') \mathbb{G}^{\mathsf{A}}_{ac}(\mathbf{r}', \mathbf{u})\mathbb{G}_{0,cb}^{-1\dagger}(\mathbf{u}, \mathbf{r})
    ]
    \label{eq:generalThetaHexpression}
\end{align}
where repeated indices are to be summed or integrated over. Integration is over all space except for $\mathbf{r}\in V_{\beta}$ and $\mathbf{r}'\in V_{\alpha}.$

\subsection{Temperature perturbation}

Consider a change in $\boldsymbol{\tau}^{(\beta)},$ the torque on $\beta,$ when all objects are at rest and at the same temperature but then the temperature of object $\alpha$ is perturbed to nonequilibrium.
We find that
\begin{align}
\frac{d\langle\boldsymbol{\tau}^{(\beta)}\rangle_{\mathrm{neq}}}{dT_{\alpha}}\bigg|_{\{T_{\alpha}\} = T_{\mathrm{eq}} = T}
    = 
    -\frac{1}{k_{B}T^{2}}\int_{0}^{\infty} \mathrm{d}t \langle \boldsymbol{\tau}^{(\beta)}(t)H^{(\alpha)}(0)\rangle_{\mathrm{eq}},
    \label{eq:lrchangetemp}
\end{align}
confirming a relationship between variations in nonequilibrium torque and the equilibrium correlation of torque and power absorption. 
The equality is established by directly computing the right-hand side and comparing to a calculation of the left-hand side starting from the trace expression for nonequilibrium Casimir torque presented in Ref.~\cite{strekha2022tracenoneq}.
Intermediate steps are outlined below. The explicit result of the correlation function in Eq.~\eqref{eq:lrchangetemp} is provided in Eq.~\eqref{eq:tauHcorrelatorresult}.
Using
\begin{align}
    \mathbb{G} &= (1 + \mathbb{G}_{0}\mathbb{T}_{\bar{1}})\frac{1}{1 - \mathbb{G}_{0}\mathbb{T}_{1}\mathbb{G}_{0}\mathbb{T}_{\bar{1}}}(1 + \mathbb{G}_{0}\mathbb{T}_{1})\mathbb{G}_{0}, \label{eq:Gexpression1}\\
    &= (1 + \mathbb{G}_{0}\mathbb{T}_{1})\frac{1}{1 - \mathbb{G}_{0}\mathbb{T}_{\bar{1}}\mathbb{G}_{0}\mathbb{T}_{1}}(1 + \mathbb{G}_{0}\mathbb{T}_{\bar{1}})\mathbb{G}_{0}.
    \label{eq:Gexpression2}
\end{align}
in Eq.~\eqref{eq:generalThetaHexpression}, we find
\begin{align}
    \int_{0}^{\infty} \mathrm{d}t ~ \langle \boldsymbol{\tau}^{(1)}(t)H^{(\alpha)}(0)\rangle_{\mathrm{eq}}
    =
    \frac{2}{\pi} \int_{0}^{\infty} \mathrm{d}\omega ~  \hbar\omega e^{\frac{\hbar\omega}{k_{B}T}}n(\omega, T)^2
    \text{Im}\text{Tr}[
    \vecop{J}\mathbb{M}_{\alpha}^{(1)}
    ]
    \label{eq:tauHcorrelatorresult}
\end{align}
where
\begin{align}
    \mathbb{M}_{1}^{(1)}
    =
    &(1 + \mathbb{G}_{0}\mathbb{T}_{2})
    \frac{1}{1-\mathbb{G}_{0}\mathbb{T}_{1}\mathbb{G}_{0}\mathbb{T}_{2}}
    \mathbb{G}_{0}
    (\mathbb{T}_{1}^{\mathsf{A}} -
    \mathbb{T}_{1}\mathbb{G}_{0}^{\mathsf{A}}\mathbb{T}_{1}^{\dagger})
    \frac{1}{1-\mathbb{G}_{0}^{\dagger}\mathbb{T}_{2}^{\dagger}\mathbb{G}_{0}^{\dagger}\mathbb{T}_{1}^{\dagger}},
\end{align}
\begin{align}
    \mathbb{M}_{2}^{(1)}
    =
    &(1 + \mathbb{G}_{0}\mathbb{T}_{1})
    \frac{1}{1-\mathbb{G}_{0}\mathbb{T}_{2}\mathbb{G}_{0}\mathbb{T}_{1}}
    \mathbb{G}_{0}
    (\mathbb{T}_{2}^{\mathsf{A}} -
    \mathbb{T}_{2}\mathbb{G}_{0}^{\mathsf{A}}\mathbb{T}_{2}^{\dagger})\mathbb{G}_{0}^{\dagger}
    \frac{1}{1-\mathbb{T}_{1}^{\dagger}\mathbb{G}_{0}^{\dagger}\mathbb{T}_{2}^{\dagger}\mathbb{G}_{0}^{\dagger}}
    \mathbb{T}_{1}^{\dagger}.
\end{align}
A key step is to note that $\mathbb{T}_{\alpha,ab}(\mathbf{u}, \mathbf{v})$ vanishes if one of spatial arguments
$\mathbf{u},\mathbf{v}$ is outside $V_{\alpha},$ which eventually allows one to extend the integrals to be over all space, leading to a trace expression.

The Casimir torque acting on an arbitrary object $\alpha$ is given by~\cite{strekha2022tracenoneq}
\begin{align}
    \langle\boldsymbol{\tau}^{(\alpha)}\rangle_{\mathrm{neq}}(T_{\mathrm{eq}}, \{T_{\beta}\}) &= \langle\boldsymbol{\tau}^{(\alpha)}\rangle_{\mathrm{eq}}(T_{\mathrm{eq}}) +
    \sum_{\beta}[\boldsymbol{\tau}_{\beta}^{(\alpha)}(T_{\beta}) - \boldsymbol{\tau}_{\beta}^{(\alpha)}(T_{\mathrm{eq}})].
    \label{eq:torqueeqplusneq}
\end{align}
That is, the total Casimir torque in nonequilibrium can be written as
a sum of an equilibrium contribution $\langle\boldsymbol{\tau}^{(\alpha)}\rangle_{\mathrm{eq}}(T_{\mathrm{eq}})$ where all objects are at a temperature $T_{\mathrm{eq}}$ plus
nonequilibrium contributions when the objects $1, \dots, N$ deviate
from the temperature of the background environment $T_{\mathrm{eq}}.$ $\boldsymbol{\tau}_{\beta}^{(\alpha)}(T)$
is the torque on $\alpha$ due to sources in $\beta,$ when body $\beta$
is at a temperature $T.$ From Eqs.~(23) and (24) in Ref.~\cite{strekha2022tracenoneq}, 
\begin{align}
    \boldsymbol{\tau}_{1,2}^{(1)}(T) &= -\frac{2}{\pi} \int_{0}^{\infty} \mathrm{d}\omega ~  n(\omega, T)
    \text{Im}\text{Tr}[
    \vecop{J}\mathbb{M}_{1,2}^{(1)}
    ].
    \label{eq:tauon1}
\end{align}
From Eqs.~\eqref{eq:torqueeqplusneq} and \eqref{eq:tauon1}, it follows that
\begin{align}
\frac{d\langle\boldsymbol{\tau}^{(1)}\rangle_{\mathrm{neq}}}{dT_{\alpha}}\bigg|_{T}
    &=
    -\frac{2}{\pi k_{B}T^{2}} \int_{0}^{\infty} \mathrm{d}\omega ~  \hbar\omega e^{\frac{\hbar\omega}{k_{B}T}}n(\omega, T)^2
    \text{Im}\text{Tr}[
    \vecop{J}\mathbb{M}_{\alpha}^{(1)}
    ],
    \label{eq:dToftauon1}
\end{align}
which combined with Eq.~\eqref{eq:tauHcorrelatorresult} proves Eq.~\eqref{eq:lrchangetemp}.

\section{Bounds on surface-parallel Casimir forces}

\begin{figure}
\centering
\includegraphics[width=0.6\columnwidth]{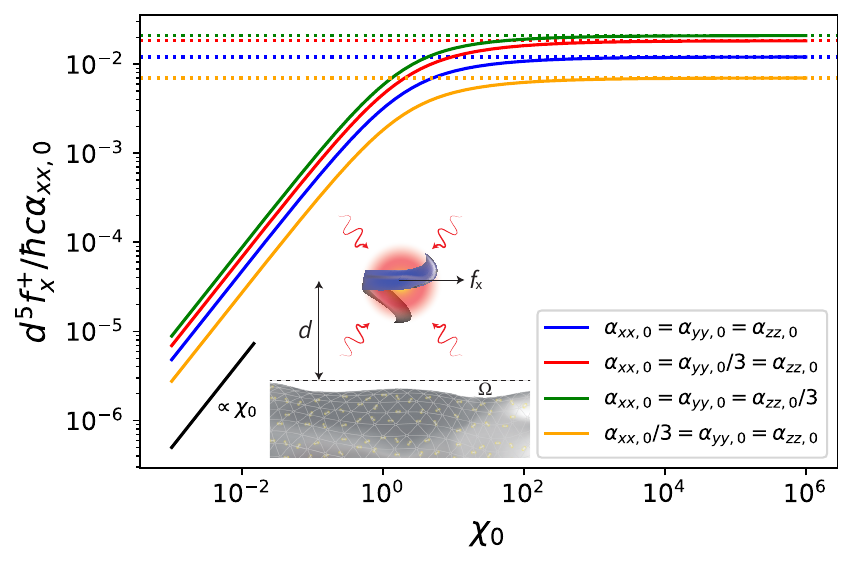}
\caption{
    \textbf{Material dependence of bounds on Casimir force}. 
    Upper bounds to the Casmir force at zero temperature along the surface-parallel direction on a nondispersive dipole above a half-space
    design domain $\Omega$ which contains a structure made of nondispersive susceptibility $\chi_{0}.$
    As $\chi_{0} \to\infty,$ the bounds approach the perfect electrical conductor limit given by Eq.~\eqref{eq:forcexchi0inflimit}, marked by horizontal dotted lines.
     }
\label{fig:forcexvschi0}
\end{figure}

Of course, similar expressions such as
\begin{align}
  (\hat{p}_{x}^{2}\Gsca)(i\xi, d\mathbf{e}_{z}, d\mathbf{e}_{z}) 
  &= 
  \frac{\hbar^{2}}{2}
  \int_{0}^{2\pi} \int_{0}^{\infty}
  (k\cos{\psi})^{2}
  (\mathbb{M}^{\mathrm{s}}(i\xi, \mathbf{k}) +
  \mathbb{M}^{\mathrm{p}}(i\xi, \mathbf{k})) 
  e^{-2\kappa_{z}d}~\frac{k\mathrm{d}k~\mathrm{d}\psi}{(2\pi)^{2}}, \\
  &=
  \frac{\hbar^{2}}{128\pi d^{5}}(9 + 18(\frac{\xi d}{c}) + 16(\frac{\xi d}{c})^2 + 8(\frac{\xi d}{c})^3)
  \exp(-2\frac{\xi d}{c})
  \begin{bmatrix}
      1 & 0 & 0 \\
      0 & 0 & 0 \\
      0 & 0 & 0
  \end{bmatrix} \nonumber \\
  &+
  \frac{\hbar^{2}}{128\pi d^{5}}(3 + 6(\frac{\xi d}{c}) + 8(\frac{\xi d}{c})^2 + 8(\frac{\xi d}{c})^3)
  \exp(-2\frac{\xi d}{c})
  \begin{bmatrix}
      0 & 0 & 0 \\
      0 & 1 & 0 \\
      0 & 0 & 0
  \end{bmatrix} \nonumber \\
  &+
  \frac{\hbar^{2}}{32\pi d^{5}}(3 + 6(\frac{\xi d}{c}) + 4(\frac{\xi d}{c})^2)
  \exp(-2\frac{\xi d}{c})
  \begin{bmatrix}
      0 & 0 & 0 \\
      0 & 0 & 0 \\
      0 & 0 & 1
  \end{bmatrix}
  \label{eq:pxpxGscaexplicit}
\end{align}
can also be used to derive bounds on lateral forces in the PEC limit on an anisotropic dipole above the half-space planar design domain.
The expression for the bound on surface-parallel forces on a dipole above a half-space design domain in the PEC limit follows 
by, without loss of generality, setting $\hat{\Theta} = \hat{p}_{x}$ resulting in
    \begin{align}
        f_{\mathrm{dip},x}^{\pm,\textrm{PEC}}(T = 0) &= 
          \pm\frac{\hbar c}{64\pi d^{5}}
          \int_{0}^{\infty}
          \alpha_{xx}(\frac{i\tilde{\xi}c}{d})
          \bigg(
          9 + 36\tilde{\xi} + 88\tilde{\xi}^2 + 112\tilde{\xi}^3 + 80\tilde{\xi}^4 + 32\tilde{\xi}^5
          \bigg)^{1/2}e^{-2\tilde{\xi}}
          ~\frac{\mathrm{d}\tilde{\xi}}{2\pi} \nonumber \\
          &
          \pm\frac{\hbar c}{64\pi d^{5}}
          \int_{0}^{\infty}
          \alpha_{yy}(\frac{i\tilde{\xi}c}{d})
          \bigg(
          3 + 12\tilde{\xi} + 32\tilde{\xi}^2 + 48\tilde{\xi}^3 + 48\tilde{\xi}^4 + 32\tilde{\xi}^5
          \bigg)^{1/2}e^{-2\tilde{\xi}}
          ~\frac{\mathrm{d}\tilde{\xi}}{2\pi} \nonumber \\
          &
          \pm\frac{\hbar c}{16\sqrt{2}\pi d^{5}}
          \int_{0}^{\infty}
          \alpha_{zz}(\frac{i\tilde{\xi}c}{d})
          \bigg(
          3 + 12\tilde{\xi} + 16\tilde{\xi}^2 + 8\tilde{\xi}^3
          \bigg)^{1/2}e^{-2\tilde{\xi}}
          ~\frac{\mathrm{d}\tilde{\xi}}{2\pi}.
          \label{eq:forcexexpandedbounds}
    \end{align}
Assuming quasistatic polarizabilities of the dipole gives
\begin{align}
    f_{\mathrm{dip},x}^{\pm,\textrm{PEC}}(T = 0) 
    &=
    \pm\bigg(
    \frac{\hbar c}{128\pi^{2} d^{5}}\gamma_{xx}\alpha_{xx,0}
    + \frac{\hbar c}{128\pi^{2} d^{5}}\gamma_{yy}\alpha_{yy,0}
    + \frac{\hbar c}{32\sqrt{2}\pi^{2} d^{5}}\gamma_{zz}\alpha_{zz,0}
    \bigg), 
    \label{eq:forcexchi0inflimit}\\
    &\approx 
    \pm\bigg(
    \frac{\hbar c}{128\pi^{2} d^{5}}5.64958\alpha_{xx,0}
    + \frac{\hbar c}{128\pi^{2} d^{5}}3.95399\alpha_{yy,0}
    + \frac{\hbar c}{32\sqrt{2}\pi^{2} d^{5}}1.98309\alpha_{zz,0}
    \bigg),
    \label{eq:forcexchi0inflimitnum}
\end{align}
where
\begin{align}
    \gamma_{xx} &\equiv
    \int_{0}^{\infty}
    \bigg(
    9 + 36\tilde{\xi} + 88\tilde{\xi}^2 + 112\tilde{\xi}^3 + 80\tilde{\xi}^4 + 32\tilde{\xi}^5
    \bigg)^{1/2}e^{-2\tilde{\xi}}
    ~\mathrm{d}\tilde{\xi}, \\
    \gamma_{yy} &\equiv
    \int_{0}^{\infty}
    \bigg(
    3 + 12\tilde{\xi} + 32\tilde{\xi}^2 + 48\tilde{\xi}^3 + 48\tilde{\xi}^4 + 32\tilde{\xi}^5
    \bigg)^{1/2}e^{-2\tilde{\xi}}
    ~\mathrm{d}\tilde{\xi}, \\
    \gamma_{zz} &\equiv
    \int_{0}^{\infty}
    \bigg(
    3 + 12\tilde{\xi} + 16\tilde{\xi}^2 + 8\tilde{\xi}^3
    \bigg)^{1/2}e^{-2\tilde{\xi}}
    ~\mathrm{d}\tilde{\xi}.
    \end{align}
See Fig.~\ref{fig:forcexvschi0} for a numerical evaluation of the bounds for finite nondispersive $\chi_{0}$ and comparison (horizontal dotted lines) to the PEC limit given by Eq.~\eqref{eq:forcexchi0inflimit}.
For finite $T$, the expression for the bound on surface-parallel forces on a dipole above a half-space design domain in the PEC limit is
    \begin{align}
        f_{\mathrm{dip},x}^{\pm,\textrm{PEC}}(T > 0) &= 
          \pm\frac{k_{B}T}{64\pi d^{4}}
          \sum_{n=0}^{\infty}{}'
          \alpha_{xx}(i\xi_{n}) 
          \bigg(
          9 + 36(\xi_{n}d/c) + 88(\xi_{n}d/c)^2 + 112(\xi_{n}d/c)^3 + 80(\xi_{n}d/c)^4 + 32(\xi_{n}d/c)^5
          \bigg)^{1/2} e^{-2\xi_{n}d/c}
          \nonumber \\
          &
          \pm\frac{k_{B}T}{64\pi d^{4}}
          \sum_{n=0}^{\infty}{}'
          \alpha_{yy}(i\xi_{n}) 
          \bigg(
          3 + 12(\xi_{n}d/c) + 32(\xi_{n}d/c)^2 + 48(\xi_{n}d/c)^3 + 48(\xi_{n}d/c)^4 + 32(\xi_{n}d/c)^5
          \bigg)^{1/2}e^{-2\xi_{n}d/c}
          \nonumber \\
          &
          \pm\frac{k_{B}T}{16\sqrt{2}\pi d^{4}}
          \sum_{n=0}^{\infty}{}'
          \alpha_{zz}(i\xi_{n}) 
          \bigg(
          3 + 12(\xi_{n}d/c) + 16(\xi_{n}d/c)^2 + 8(\xi_{n}d/c)^3
          \bigg)^{1/2}e^{-2\xi_{n}d/c}
          \label{eq:forcexexpandedboundsT}
    \end{align}
where $\xi_{n} = 2\pi k_{\mathrm{B}} Tn/\hbar$, and the prime on the
summation implies a prefactor of $1/2$ for the contribution at $n =
0$.

\section{Bianisotropy}
The prior scattering operator formalism worked in
terms of the Green's function of the system $\mathbb{G}$, defined by
$\left[ \nabla\times\nabla\times - \mathbb{V} -
  \frac{\omega^{2}}{c^{2}}\mathbb{I}\right]\mathbb{G}(\mathbf{r},
\mathbf{r}') = \mathbb{I}\delta^{(3)}(\mathbf{r}, \mathbf{r}')$ where
$\mathbb{V} = \frac{\omega^{2}}{c^{2}}(\mathbb{\epsilon} - \mathbb{I})
+ \nabla\times(\mathbb{I} - \mathbb{\mu}^{-1})\nabla\times$ is the
potential or generalized susceptibility introduced by the
objects~\cite{kruger_nonequilbrium_fluctuations_review,strekha2022tracenoneq}.
While this form of $\mathbb{V}$ can account for polarization and magnetization currents ($
    \mathbb{V}\mathbf{E} = i\mu_{0}\omega(-i\omega\mathbf{P} + \nabla\times\mathbf{M}))
$
one can work in a 6-vector notation to explicitly separate the electric and magnetic dipolar responses so as to more easily extend formalism to bianisotropic systems.

\subsection{Dipole convention}

We consider bianisotropic particles whose electric and magnetic dipole moments respond linearly to $\mathbf{E}, \mathbf{B}$ fields:
\begin{align}
    \mathbf{p} &= \alpha_{ee}\mathbf{E} + \alpha_{em}\mathbf{B} \\
    \mathbf{m} &= \alpha_{me}\mathbf{E} + \alpha_{mm}\mathbf{B}.
\end{align}
In order to obtain the correlation functions between the dipoles from the FDT, it is convenient to rescale variables so that everything is in the same units. Therefore, define a dipole $\mathbf{d}_{e} = \mathbf{p}$ and $\mathbf{d}_{m} = \mathbf{m}/c,$ where $c$ is the speed of light and fields $\mathbf{F}_{e} = \mathbf{E}$ and $\mathbf{F}_{m} = c\mathbf{B}$ so that
\begin{align}
    \mathbf{d}_{e} &= \epsilon_{0}\tilde{\alpha}_{ee}\mathbf{F}_{e} + \epsilon_{0}\tilde{\alpha}_{em}\mathbf{F}_{m} \\
    \mathbf{d}_{m} &= \epsilon_{0}\tilde{\alpha}_{me}\mathbf{F}_{e} + \epsilon_{0}\tilde{\alpha}_{mm}\mathbf{F}_{m}
\end{align}
where the polarizability matrices also need to be rescaled so that
\begin{align}
    \begin{bmatrix}
    \tilde{\alpha}_{ee} & \tilde{\alpha}_{em} \\
    \tilde{\alpha}_{me} & \tilde{\alpha}_{mm}
    \end{bmatrix}
    \equiv
    \frac{1}{\epsilon_{0}}\begin{bmatrix}
    \alpha_{ee} & \frac{1}{c}\alpha_{em} \\
    \frac{1}{c}\alpha_{me} & \frac{1}{c^{2}}\alpha_{mm}
    \end{bmatrix}.
\end{align}
This is done so that $\mathbf{d}$ has the SI units of electric dipoles (C$\cdot$m) and the polarizabilities have units of volume (m$^{3}$). We follow the convention where
\begin{align}
    \tilde{\alpha}_{ee}(\omega) = 4\pi R^{3}\frac{\MatrixVariable{\epsilon}(\omega) - \mathbb{I}}{\MatrixVariable{\epsilon}(\omega) + 2\mathbb{I}},
\end{align}
i.e. a factor of $4\pi$ is included.

\subsection{Green's function conventions}

The electric and magnetic fields generated by the dipole is given by
\begin{align}
    \mathbf{E} &= \frac{k^{2}}{\epsilon_{0}}[\mathbb{G}_{0}^{ee}\mathbf{d}_{e} + \mathbb{G}_{0}^{em}\mathbf{d}_{m}], \\
    \mathbf{B} &= \frac{k^{2}}{c\epsilon_{0}}[\mathbb{G}_{0}^{me}\mathbf{d}_{e} + \mathbb{G}_{0}^{mm}\mathbf{d}_{m}],
\end{align}
where
\begin{align}
    \mathbb{G}_{0}^{ee} &= \mathbb{G}_{0}^{mm} =  \frac{e^{ikr}}{4\pi r}\left[a\mathbb{I} + b \mathbf{e}_{r}\otimes\mathbf{e}_{r}\right], \\
    \mathbb{G}_{0}^{me} &= - \mathbb{G}_{0}^{em} = \frac{e^{ikr}}{4\pi r}l\left[\mathbf{e}_{\phi}\otimes\mathbf{e}_{\theta} - \mathbf{e}_{\theta}\otimes\mathbf{e}_{\phi}\right],
\end{align}
where $k = \omega/c,$ 
\begin{align}
    a &= 1 + \frac{ikr - 1}{(kr)^{2}}, \\
    b &= \frac{3 - 3ikr - (kr)^{2}}{(kr)^{2}}, \\
    l &= 1 + \frac{i}{kr},
\end{align}
and $\mathbf{e}_{r} = (\mathbf{R} - \mathbf{R}')/|\mathbf{R} - \mathbf{R}'|$ in a unit vector from the location of the dipole $\mathbf{R}'$ to the observation point $\mathbf{R}.$
Defining
\begin{align}
    \mathbb{G}_{0}(\mathbf{r}, \mathbf{r}') =
    \begin{bmatrix}
    \mathbb{G}^{ee}_{0}(\mathbf{r}, \mathbf{r}') & \mathbb{G}^{em}_{0}(\mathbf{r}, \mathbf{r}') \\
    \mathbb{G}^{me}_{0}(\mathbf{r}, \mathbf{r}') & \mathbb{G}^{mm}_{0}(\mathbf{r}, \mathbf{r}')
    \end{bmatrix},
\end{align}
we have
\begin{align}
\mathbb{G}_{0}^{\mathsf{A}}(\mathbf{r}, \mathbf{r}) = \frac{k}{6\pi}\mathbb{I},
\end{align}
where $\mathbb{I}$ is a 6-by-6 identity matrix in this case. Also,
\begin{align}
    \mathbf{F} = \mu_{0}\omega^{2}\mathbb{G}_{0}\mathbf{d}.
\end{align}

\subsection{Trace expressions in terms of scattering operators}

Next, we derive the general expressions for heat absorption, force, and torque on a body with volume $V$ assuming electric and magnetic dipolar responses.
The expressions are in the Fourier frequency space and we only consider 
one frequency integral here in order to simplify the expressions, since ultimately we will perform
an ensemble average where the different frequency components are uncorrelated, according to the fluctuation-dissipation theorem~\cite{novotny2012principles}.

The energy absorbed by an object is given by
\begin{align}
    H &= \int_{-\infty}^{\infty}\frac{\mathrm{d}\omega}{2\pi}
    \int_{V}\mathrm{d}^{3}\mathbf{r}\mathbf{J}_{tot}^{*}(\omega,\mathbf{r})\cdot\mathbf{E}_{tot}(\omega,\mathbf{r})
    \\
    &=  \int_{-\infty}^{\infty}\frac{\mathrm{d}\omega}{2\pi}
    \int_{V}\mathrm{d}^{3}\mathbf{r}[ i\omega\mathbf{P}^{*}(\omega,\mathbf{r})  + \nabla\times\mathbf{M}^{*}(\omega,\mathbf{r})]\cdot\mathbf{E}_{tot}(\omega,\mathbf{r}) \\
    &= -\int_{-\infty}^{\infty}\frac{\mathrm{d}\omega}{2\pi}
    \int_{V}\mathrm{d}^{3}\mathbf{r} ~\omega~\textrm{Im}[\mathbf{P}^{*}(\omega,\mathbf{r})\cdot\mathbf{E}_{tot}(\omega,\mathbf{r}) + \mathbf{M}^{*}(\omega,\mathbf{r})\cdot\mathbf{B}_{tot}(\omega,\mathbf{r})] \\
    &\equiv -\int_{-\infty}^{\infty}\frac{\mathrm{d}\omega}{2\pi}
    \int_{V}\mathrm{d}^{3}\mathbf{r} ~\omega~\textrm{Im}[\mathbf{D}^{*}(\omega,\mathbf{r})\cdot\mathbf{F}_{tot}(\omega,\mathbf{r})] \\
    &\equiv -\int_{-\infty}^{\infty}\frac{\mathrm{d}\omega}{2\pi}
    \int_{V}\mathrm{d}^{3}\mathbf{r} ~\frac{1}{\hbar}~\textrm{Im}[\mathbf{D}^{*}(\omega,\mathbf{r})\Theta_{E}\mathbf{F}_{tot}(\omega,\mathbf{r})]
\end{align}
where we defined a dipole density field $\mathbf{D}(\omega,\mathbf{r})$ (not to be confused with the displacement field) and a 6-vector electromagnetic field $\mathbf{F}(\omega, \mathbf{r}):$
\begin{align}
    \mathbf{D}(\omega,\mathbf{r}) &=
    \begin{bmatrix}
    \mathbf{P}(\omega,\mathbf{r}) \\
    \mathbf{M}(\omega,\mathbf{r})/c
    \end{bmatrix}, \\
    \mathbf{F}(\omega,\mathbf{r}) &=
    \begin{bmatrix}
    \mathbf{E}(\omega,\mathbf{r}) \\
    \mathbf{B}(\omega,\mathbf{r})c
    \end{bmatrix}.
\end{align}
Given a dipole distribution $\mathbf{D}(\omega,\mathbf{r}),$ the field $\mathbf{F}(\omega,\mathbf{r})$ is given by $\mathbf{F} = \mu_{0}\omega^{2}\mathbb{G}_{0}\mathbf{D}.$ In deriving the expressions above, we made use of $\mathbf{J}_{tot}(t,\mathbf{r}) = \mathbf{J}_{f}(t,\mathbf{r}) + \frac{\partial \mathbf{P}(t,\mathbf{r})}{\partial t} + \nabla\times\mathbf{M}(t,\mathbf{r})$ with the free current $\mathbf{J}_{f}(t,\mathbf{r})$ set to zero, Faraday's law $\nabla\times\mathbf{E}(t,\mathbf{r}) = -\frac{\partial\mathbf{B}(t,\mathbf{r})}{\partial t},$ and the vector identity
\begin{align}
    \nabla\cdot(\mathbf{M}^{*}\times\mathbf{E}) = (\nabla\times\mathbf{M})^{*}\cdot\mathbf{E} - \mathbf{M}^{*}\cdot(\nabla\times\mathbf{E})
\end{align}
to throw away a surface term since the magnetization right outside the body is assumed to be 0. $\Theta_{E} \equiv \hbar\omega\mathbb{I}$ is an energy operator defined which in a Cartesian 6-by-6 Dyadic form satisfies $\Theta_{E}(\mathbf{x}, \mathbf{y}) = \delta^{(3)}(\mathbf{x} - \mathbf{y}) \hbar\omega \mathbb{I}_{6\times 6}.$

Using similar steps, one can show (discarding vanishing surface terms) that the force $\mathbf{f}$ is equal to
\begin{align}
    \mathbf{f} &= \int_{V} \mathrm{d}^{3}\mathbf{r}\int_{-\infty}^{\infty} \frac{\mathrm{d}\omega}{2\pi} \left[ \rho^{*}(\omega, \mathbf{r}) \mathbf{E}(\omega, \mathbf{r}) + \mathbf{J}^{*}(\omega, \mathbf{r})\times\mathbf{B}(\omega,\mathbf{r}) \right] \\
    &= \int_{V} \mathrm{d}^{3}\mathbf{r}\int_{-\infty}^{\infty} \mathrm{d}\omega\frac{1}{\hbar\omega}J_{a}^{*} \vecop{p} E_{a} \\
    &= \int_{V} \mathrm{d}^{3}\mathbf{r}\int_{-\infty}^{\infty} \frac{\mathrm{d}\omega}{2\pi}\frac{1}{\hbar\omega}i\omega(P_{a}^{*} \vecop{p} E_{a} + M_{a}^{*} \vecop{p} B_{a}) \\
    &= \int_{V} \mathrm{d}^{3}\mathbf{r}\int_{-\infty}^{\infty} \frac{\mathrm{d}\omega}{2\pi}\frac{1}{\hbar\omega}i\omega D_{a}^{*} \vecop{p}F_{a} \\
    &= -\int_{V} \mathrm{d}^{3}\mathbf{r}\int_{-\infty}^{\infty} \frac{\mathrm{d}\omega}{2\pi}\frac{1}{\hbar}\text{Im}[\mathbf{D}^{*} \Theta_{\mathbf{p}}\mathbf{F}]
    \label{eq:compactforceintegral}
\end{align}
where $\Theta_{\mathbf{p}}$ is a linear momentum operator which in a Cartesian 6-by-6 Dyadic form satisfies 
\begin{align}
    \Theta_{\mathbf{p}}(\mathbf{x}, \mathbf{y}) = \mathbf{e}_{a}\delta^{(3)}(\mathbf{x} - \mathbf{y}) \sum_{a = x,y,z}
    \begin{bmatrix}
    \hat{p}_{a} & 0 & 0 & 0 & 0 & 0 \\
    0 & \hat{p}_{a} & 0 & 0 & 0 & 0 \\
    0 & 0 & \hat{p}_{a} & 0 & 0 & 0 \\
    0 & 0 & 0 & \hat{p}_{a} & 0 & 0 \\
    0 & 0 & 0 & 0 & \hat{p}_{a} & 0 \\
    0 & 0 & 0 & 0 & 0 & \hat{p}_{a}
    \end{bmatrix}
\end{align}
where $\hat{\mathbf{p}} = -i\hbar\nabla$ is the momentum operator.

The torque $\boldsymbol{\tau}$ (discarding vanishing surface terms) is
\begin{align}
    \boldsymbol{\tau}
    &= \int_{V} \mathrm{d}^{3}\mathbf{r}\int_{-\infty}^{\infty} \frac{\mathrm{d}\omega}{2\pi} \mathbf{r}\times [ \rho^{*}(\omega, \mathbf{r}) \mathbf{E}(\omega, \mathbf{r}) + \mathbf{J}^{*}(\omega, \mathbf{r})\times\mathbf{B}(\omega,\mathbf{r}) ] \\
    &= \int_{V} \mathrm{d}^{3}\mathbf{r}\int_{-\infty}^{\infty} \frac{\mathrm{d}\omega}{2\pi}
    \frac{1}{\hbar\omega}[ J_{a}^{*} \vecop{L} E_{a} +  \mathbf{J}^{*}\vecop{S}\mathbf{E}] \\
    &= \int_{V} \mathrm{d}^{3}\mathbf{r}\int_{-\infty}^{\infty} \frac{\mathrm{d}\omega}{2\pi}
    \frac{1}{\hbar\omega}i\omega[P_{a}^{*} \vecop{L} E_{a} +  \mathbf{P}^{*}\vecop{S}\mathbf{E} + M_{a}^{*} \vecop{L} B_{a} +  \mathbf{M}^{*}\vecop{S}\mathbf{B}] \\
    &= \int_{V} \mathrm{d}^{3}\mathbf{r}\int_{-\infty}^{\infty} \frac{\mathrm{d}\omega}{2\pi}
    \frac{1}{\hbar\omega}i\omega[D_{a}^{*} \vecop{L} F_{a} +  \mathbf{D}^{*}\vecop{S}\mathbf{F}] \\
    &= -\int_{V} \mathrm{d}^{3}\mathbf{r}\int_{-\infty}^{\infty} \frac{\mathrm{d}\omega}{2\pi}
    \frac{1}{\hbar}\text{Im}[\mathbf{D}^{*}\Theta_{\mathbf{J}}\mathbf{F}]
\end{align}
where $\Theta_{\mathbf{J}}$ is a total angular momentum operator
\begin{align}
    \Theta_{\mathbf{J}}(\mathbf{x}, \mathbf{y}) = \mathbf{e}_{a}\delta^{(3)}(\mathbf{x} - \mathbf{y}) \sum_{a = x,y,z}
    \begin{bmatrix}
    \hat{L}_{a}\mathbb{I}_{3\times 3} + \hat{S}_{a} &  0_{3\times 3} \\
    0_{3\times 3} & \hat{L}_{a}\mathbb{I}_{3\times 3} + \hat{S}_{a}
    \end{bmatrix}.
\end{align}
$\hat{\mathbf{L}} = \mathbf{r}\times\hat{\mathbf{p}} = -i\hbar\mathbf{r}\times\nabla$ is the orbital angular momentum operator and, in the Cartesian basis, $(\hat{S}_{a})_{bc} = -i\hbar
\epsilon_{abc}$ is the spin angular momentum operator.

In a nonequilibrium stationary state, each object is assumed to be at local
equilibrium, such that the current fluctuations within each object
satisfy the FDT at the appropriate local temperature. The details of
the use of the FDT and local equilibrium properties with the scattering
equations have been described
before~\cite{kruger_nonequilbrium_fluctuations_review,kruger_trace_formulae_for_nonequilibrium,strekha2022tracenoneq}.
Following similar steps leads to the follow expression for the nonequilibrium expressions
\begin{align}
    H_{\textrm{body}} &= \int_{0}^{\infty}\mathrm{d}\omega [n(\omega, T_{\textrm{body}}) - n(\omega, T_{\textrm{env}})] \frac{2}{\pi}\textrm{Im}\text{Tr}[(-\Theta_{E}\mathbb{G}_{0})(\mathbb{T}^{\mathsf{A}} - \mathbb{T}\mathbb{G}_{0}^{\mathsf{A}}\mathbb{T}^{\dagger})], \\
    &= \int_{0}^{\infty}\mathrm{d}\omega [n(\omega, T_{\textrm{body}}) - n(\omega, T_{\textrm{env}})] \frac{2}{\pi}\text{Tr}[(-\Theta_{E}\mathbb{G}_{0}^{\mathsf{A}})(\mathbb{T}^{\mathsf{A}} - \mathbb{T}\mathbb{G}_{0}^{\mathsf{A}}\mathbb{T}^{\dagger})],
\end{align}
\begin{align}
    \mathbf{f}_{\textrm{body}} &= \int_{0}^{\infty}\mathrm{d}\omega [n(\omega, T_{\textrm{body}}) - n(\omega, T_{\textrm{env}})] \frac{2}{\pi}\textrm{Im}\text{Tr}[(-\Theta_{\mathbf{p}}\mathbb{G}_{0})(\mathbb{T}^{\mathsf{A}} - \mathbb{T}\mathbb{G}_{0}^{\mathsf{A}}\mathbb{T}^{\dagger})], \\
    &= \int_{0}^{\infty}\mathrm{d}\omega [n(\omega, T_{\textrm{body}}) - n(\omega, T_{\textrm{env}})] \frac{2}{\pi}\text{Tr}[(-\Theta_{\mathbf{p}}\mathbb{G}_{0}^{\mathsf{A}})(\mathbb{T}^{\mathsf{A}} - \mathbb{T}\mathbb{G}_{0}^{\mathsf{A}}\mathbb{T}^{\dagger})],
\end{align}
\begin{align}
    \boldsymbol{\tau}_{\textrm{body}} &= \int_{0}^{\infty}\mathrm{d}\omega [n(\omega, T_{\textrm{body}}) - n(\omega, T_{\textrm{env}})] \frac{2}{\pi}\textrm{Im}\text{Tr}[(-\Theta_{\mathbf{J}}\mathbb{G}_{0})(\mathbb{T}^{\mathsf{A}} - \mathbb{T}\mathbb{G}_{0}^{\mathsf{A}}\mathbb{T}^{\dagger})], \\
    &= \int_{0}^{\infty}\mathrm{d}\omega [n(\omega, T_{\textrm{body}}) - n(\omega, T_{\textrm{env}})] \frac{2}{\pi}\text{Tr}[(-\Theta_{\mathbf{J}}\mathbb{G}_{0}^{\mathsf{A}})(\mathbb{T}^{\mathsf{A}} - \mathbb{T}\mathbb{G}_{0}^{\mathsf{A}}\mathbb{T}^{\dagger})],
\end{align}
where the second lines for the force and torque follow if $\mathbb{G}_{0}$ is the Green's function of a translationally and rotationally invariant background, respectively.
For an isolated object in the dipolar limit, since $\mathbb{G}_{0}^{\mathsf{A}}(\mathbf{r},\mathbf{r}) = \frac{k}{6\pi}\mathbb{I}_{6\times 6},$ it follows that only the electric-electric and magnetic-magnetic subblocks of $\mathbb{T}$ contribute to far-field radiation (or force or torque), as expected. In the equilibrium case, similar expressions as in the main text hold but of course now in a 6-vector interpretation.

\bibliography{refs}